\documentclass{aa}  
\usepackage{graphicx}
\usepackage{txfonts}
\usepackage{hyperref}
\hypersetup{
    colorlinks=true,
    linkcolor=blue,
    filecolor=magenta,      
    urlcolor=cyan,
    citecolor=blue
    }
\usepackage{amsmath}
\usepackage[version=4]{mhchem}
\usepackage{gensymb}
\usepackage{chemfig}
\usepackage{comment}

\linenumbers
\renewcommand\makeLineNumber{}
    
\begin{document}

   \title{A tunable Monte Carlo method for mixing correlated-k opacities}

   \subtitle{PRAS: polynomial reconstruction and sampling}

   \author{Elspeth K.H. Lee}

   \institute{Center for Space and Habitability, University of Bern, Gesellschaftsstrasse 6, CH-3012 Bern, Switzerland}

    \date{Received DD MM YYYY / Accepted DD MM YYYY}
 
  \abstract
   {Mixed-gas opacities are critical for radiative transfer in stellar and substellar atmospheres. 
   Several approaches exist to obtain net k-coefficients for arbitrary mixtures, each trading accuracy against computational cost.}
   {I introduce a tunable Polynomial (or spline) Reconstruction And Sampling (PRAS) method to compute randomly overlapped opacities within a wavelength band.} 
   {For each species and band, PRAS fits the opacity cumulative distribution function (CDF) with a polynomial or spline, then performs a Monte Carlo convolution to form the mixed distribution. 
   A tunable trade-off between accuracy and speed of computation is controlled by the quality of the CDF fit and the total number of random samples used in the Monte Carlo integration scheme.}
   {PRAS is typically as accurate as, or more accurate, than other methods at recovering individual, pre-mixed k-coefficients with the random overlap assumption.
   In an exoplanet atmosphere outgoing spectral flux comparison test, PRAS, even with a small number of samples (250), is at worse within $\lesssim$20\% of the pre-mixed (PM) reference and typically within $\approx$5\% of the resorting and rebinning method (RORR).
   In the vertical flux and heating rate tests, PRAS produces similar results to RORR, and an improvement over the adaptive equivalent extinction (AEE) method.}
   {In the limit of exact CDF representation and infinite samples, PRAS converges to the exact, convolved randomly overlapped opacity distribution. 
   Given its accuracy and scalability to larger quadrature sets at comparable cost to RORR, PRAS is a practical alternative for retrievals and post-processing applications.}


   \maketitle

\section{Introduction}

The application of radiative transfer (RT) techniques to model substellar atmospheres is a fundamental cornerstone of exoplanet and brown dwarf research.
The utilisation of Opacity Distribution Functions (ODFs) or opacity Cumulative Distribution Functions (CDFs) in RT models is widespread between fields, from stellar atmospheres, planetary science, Earth science and exoplanet atmospheres.

The Opacity Distribution Function (ODF) method attempts to use the properties of the inverse cumulative distribution function (CDF) of the opacity in a wavelength band, rather than perform a, typically more computationally expensive, line-by-line (LBL) or opacity sampling (OS) RT calculation.
Detailed wavelength information is lost during this process, transforming the wavelength axis to a normalised cumulative weighting $g$ $\in$ [0,1].
The opacity is therefore more appropriately described and discretised in quantiles, with associated values and weights.

Opacity ODFs have been used to model stellar atmosphere for several decades since \citet{Labs1951} and \citet{Kurucz1979} detailed their implementation and use in computing stellar temperature-pressure (T-p) structures.
Typically, these classic models used pre-mixed (PM) ODF tables, taking the mixing ratio of species from chemical equilibrium (CE) calculations.
For these hotter atmospheres, the CE assumption was typically accurate enough for tackling the scientific questions at the time.
Therefore, in parameter regimes where the ODF method remains accurate or where computational efficiency is required, such as 3D atmospheric modelling, the ODF method continues to be a staple component of stellar atmosphere modelling \citep[e.g.][]{Castelli_2003, Vogler_2005, Kostogryz_2023, Stein_2024}.
However, as the need for non-equilibrium chemistry increased, in particular for cooler M-dwarf and brown dwarf atmospheres, more flexible methods were required to capture the variable opacity of the atmosphere.
\citet{Saxner1984} initially developed a flexible ODF mixing scheme for the MARCS stellar model \citep[e.g.][]{Gustafsson2008}.
This method fit a monotonic piecewise polynomial approximation to the opacity CDF, and assumed a random overlap (convolution) of lines when performing the mixing calculation.
However, this ODF mixing approach was found to be inadequate in several cases, for example, in cool, high C/O ratio stellar atmospheres \citep[e.g.][]{Ekberg_1986}, with the opacity sampling (OS) approach generally used in contemporary MARCS modelling efforts \citep{Gustafsson2008, Jorgensen2024}.

Classically, for modelling Earth and planetary atmospheres, transmission function band averaged methods \citep{Goody1952, Malkmus1967} were in use for decades.
The most utilised was the \citet{Malkmus1967} method, which assumed an uncorrelated (random) distribution of lines in a wavelength band.
Tables of Malkmus band averaged parameters ($\alpha$ and $\beta$) were required for different species at different T-p and path lengths for each band.
However, these models were inflexible when modelling inhomogeneous paths where the composition and thermal properties of the atmosphere were changing significantly.
Detailed descriptions of these classic methods can be found in \citet{Pierrehumbert2010}.

Critical to many contemporary RT calculations in substellar (brown dwarf and exoplanet) atmospheric science is the use of the correlated-k assumption, either for producing synthetic transmission, albedo and emission spectra from fluxes escaping the atmosphere \citep[e.g.][]{Barstow2020} or for calculating internal atmospheric heating rates for one-dimensional models \citep[e.g.][]{Marley2021} and three-dimensional general circulation models (GCMs) \citep[e.g.][]{Showman2009}.

Detailed explanations and examples of the fundamentals of the correlated-k assumption are available from many sources \citep[e.g.][]{Goody1989,Lacis1991,Fu1992,Irwin2008,Pierrehumbert2010,Grimm2015,Amundsen2017, Garland2019, Hogan2020}.
The correlated-k assumption is that each quantile in the opacity CDF for each band is correlated across the atmosphere, meaning, in practice, the RT calculation is performed using the same opacity quantile at each layer.
Through performing Gaussian quadrature within the RT model for each quantile and weight, an integrated band averaged flux through the atmosphere can be calculated with the k-coefficients.
This assumption allows inhomogeneous paths to be modelled in a simple manner.

Despite this correlated assumption, well-designed ODF and correlated-k frameworks have been shown to reproduce line-by-line calculations well, to within 5\% \citep[e.g.][]{Irwin2008, Garland2019, Molliere2019, Leconte2021}.
Utilising correlated-k is a standard practice in the field of substellar astrophysics, from retrieval modelling \citep[e.g.][]{Molliere2019, Barstow2020, Min2020, Barstow2022, MacDonald2023}, 1D RCE modelling \citep[e.g.][]{Malik2017, Batalha2019, Marley2021, Mukherjee2023}, and 3D general circulation models (GCMs) \citep[e.g.][]{Showman2009, Kataria2013, Amundsen2014, Lee2021, Deitrick2022, Schneider2022}.
Models vary in their individual implementation, such as the binning scheme or opacity line list source, but all retain the fundamental correlated-k assumption throughout the RT calculation.
ODF methods have also recently been used in stellar spectral synthesis codes \citep[e.g.][]{Cerentic2019, Anusha2021}.

However, despite widespread adoption of the correlated-k methodology, particular care and further assumptions must be taken when mixing individual k-tables from different species.
In this work, I propose a Monte Carlo-based random-overlap method, polynomial reconstruction and sampling (PRAS), that seeks to retain the accuracy of RORR with tunable computational cost.
In Section \ref{sec:overview}, I provide a brief overview of current k-table mixing methodologies.
In Section \ref{sec:PRAS_meth}, the PRAS method is introduced in detail and examples given of the overall approach.
In Section \ref{sec:acc}, I test the accuracy of the PRAS method for recovering k-coefficients, producing emission spectra and atmospheric heating rates.
In Section \ref{sec:speed}, I perform an initial test on the computational efficiency of the PRAS method.
Sections \ref{sec:disc} and \ref{sec:conc} contain the discussion and conclusions of this study respectively.

\section{Overview of common mixing methods}
\label{sec:overview}

When designing a correlated-k framework, a practical problem arises about the most appropriate way to mix the k-coefficients of different species together, weighted by their local volume mixing ratios (VMRs).
In effect, for input to a correlated-k RT scheme, one requires the net k-coefficients from multiple species, each contributing their individual opacities and VMRs.
Existing methods, such as pre-mixed (PM), random overlap with resorting and rebinning (RORR), and adaptive equivalent extinction (AEE), each involve trade-offs between flexibility, accuracy and computational cost. 
In this section, I briefly review common approaches for mixing k-tables in the substellar atmosphere literature. 
\citet{Amundsen2017} provide an in-depth explanation of each of these methods.

\subsection{Pre-mixed (PM)}

In the pre-mixed (PM) method, the VMRs of each species are known in advance at each T-p point, typically from pre-calculated chemical equilibrium calculations.
K-tables can then be made from the cross-section data weighted by the VMR of each species for a two-dimensional set of T-p points.
These tables can then be later interpolated to the T-p conditions present in each of the atmospheric layers.
PM tables are typically very accurate as no further assumptions are made beyond the correlated-k assumption, although care must be taken that interpolation errors do not affect the accuracy of the RT calculation \citep[e.g.][]{Amundsen2017}.
However, PM tables are inflexible in regards to the VMR of each species, making them unsuitable for calculations where the VMR is changing in a non-equilibrium fashion, such as in thermokinetic or photochemical modelling.
Some examples of substellar atmosphere models that have used PM opacities are \citet{Showman2009, Kataria2013, Baudino2015, Lee2021, Marley2021} and \citet{Mukherjee2023}.

\subsection{Random overlap with resorting and rebinning (RORR)}

The random overlap (RO) method is a method of mixing two or more k-tables weighted by their respective VMRs, 
assuming zero correlation (in wavelength) between lines of different species; in effect, assuming that there is an equal chance that lines overlap or do not overlap \citep{Goody1989, Lacis1991, Fu1992}.
However, in reality it is likely that there is some correlation, or even high correlation, between the lines of each species at certain wavelengths.
In the random overlap assumption, one therefore assumes that, overall, the opacity mixture is well represented by a random convolution of the CDF of all species.

A naive implementation of the RO method is generally far too computationally expensive to use in practice, but retains excellent accuracy compared to LBL calculations \citep{Amundsen2017}, therefore the more efficient random overlap with resorting and rebinning (RORR) method \citep{Lacis1991} has been generally employed in the substellar atmosphere literature.
The RORR method mixes two gases assuming random overlap, then reassembles the cumulative distribution function of the mixed gases to find the mixed k-coefficients.
These mixed k-coefficients are then mixed with the next gas in the list, and so on, cascading through each gas until the net k-coefficients are found.
RORR significantly speeds up the computation of random overlapped opacities compared to a naive RO implementation at the cost of some accuracy \citep{Amundsen2017}.
Various models have utilised RORR, typically retrieval and non-equilibrium chemical modelling where the VMR of different species is dynamically changing in advance of the RT computation \citep[e.g.][]{Irwin2008, Molliere2019, Phillips2020, Mukherjee2024, Line2025}.

\subsection{Adaptive equivalent extinction (AEE)}

Due to the computational expense of the RORR method, \citet{Amundsen2017} propose the adaptive equivalent extinction (AEE) method, based on the equivalent extinction (EE) method from \citet{Edwards1996}.
This method finds the species that is the most significant absorber up to an optical depth of $\tau$ $\sim$ 1, this species is then used as the template species, after which the grey opacity of all other species are added to the k-coefficients of the template species at each layer, weighted by their VMR.
The grey opacity of a species, $\sigma_{\rm grey}$ [cm$^{2}$ molecule$^{-1}$],  is given by the quadrature of the k-coefficients
\begin{equation}
    \sigma_{\rm grey} = \chi_{s}\sum_{g=1}^{N_{g}}\sigma_{g}w_{g},
\end{equation}
where $\sigma_{g}$ [cm$^{2}$ molecule$^{-1}$] are the cross section k-coefficients, $w_{g}$ the quadrature weight and $\chi_{s}$ the VMR of the species.
This method is fast and accurate enough for GCM usage \citep{Amundsen2017, Schneider2024} and it, and variants, have been utilised in exoplanet GCMs that evolve kinetic chemistry time-dependently to investigate the feedback effects of changing chemical compositions of the atmosphere \citep[e.g.][]{Drummond2020, Zamyatina2024, Lee2024}.

\section{Polynomial reconstruction and sampling (PRAS) method}
\label{sec:PRAS_meth}

\begin{figure*}
    \centering
    \includegraphics[width=0.48\linewidth]{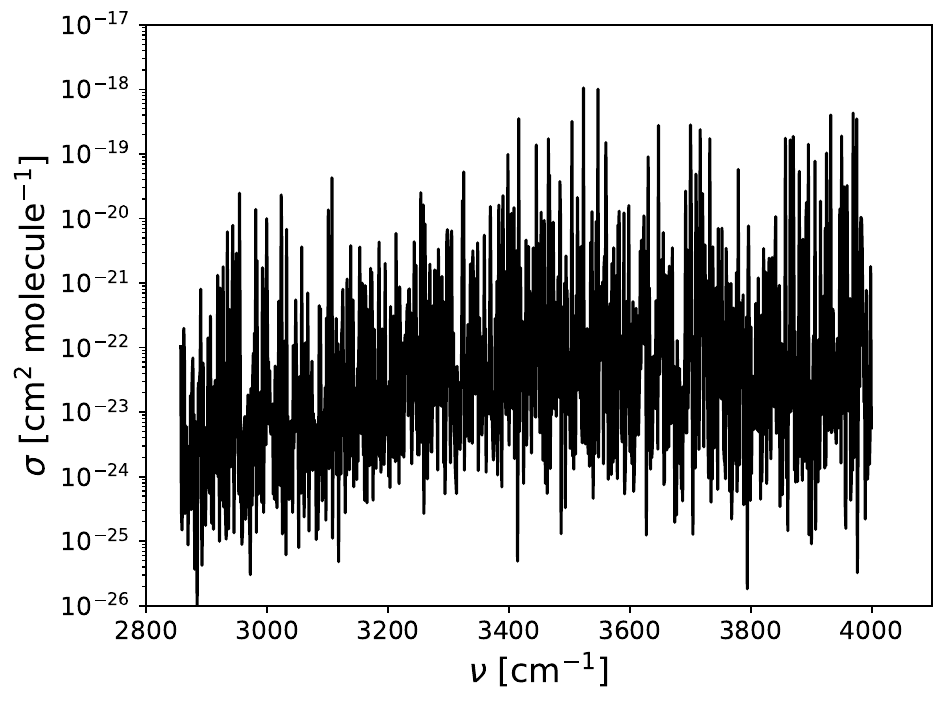}
    \includegraphics[width=0.48\linewidth]{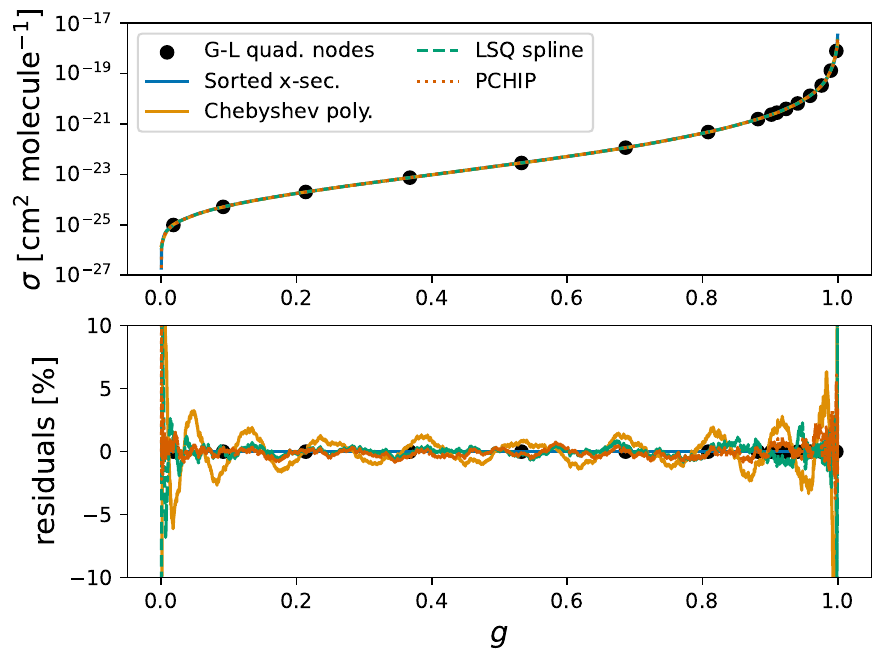}
    \caption{Illustrative example of creating a polynomial table (p-table) from cross-section data. 
    Left: \ce{H2O} cross sections from the DACE database at a temperature of 1000 K and pressure of 10$^{-8}$ bar between a wavelength range of 4.4 $\mu$m and 8.7$\mu$m, a typical hot Jupiter GCM band \citep{Kataria2013}.
    Right: approximation of the sorted cross sections using Chebyshev polynomials (20 nodes), least squares (LSQ) B-spline (32 knots), and PCHIP method (32 knots), as well as the residual error.
    Representative Legendre quadrature nodes (8+8 split) used in the petitRADTRANS radiation code \citep{Molliere2019} are shown as the black points, these values would be assigned weights and be taken as the k-coefficients in the production of k-tables.}
    \label{fig:cs_fit}
\end{figure*}

\begin{figure*}
    \centering
    \includegraphics[width=0.48\linewidth]{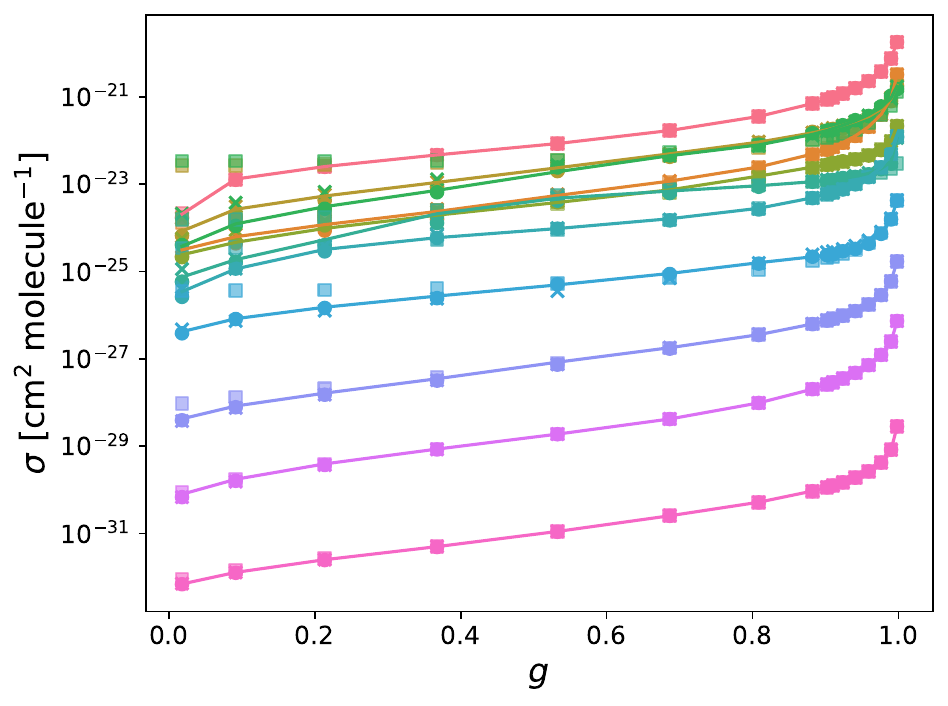}
    \includegraphics[width=0.48\linewidth]{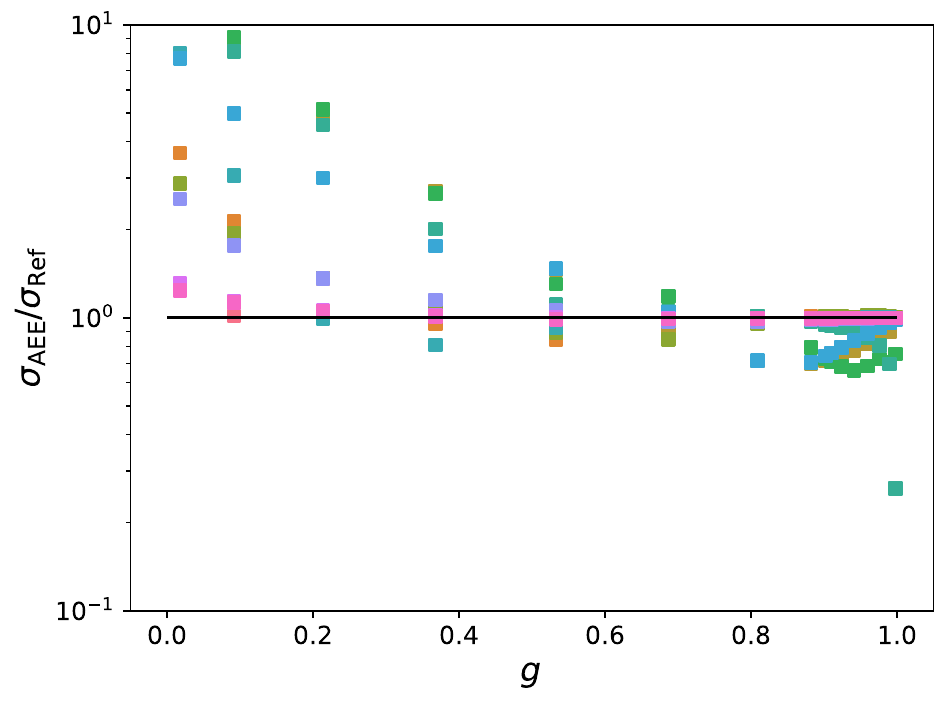}
    \includegraphics[width=0.48\linewidth]{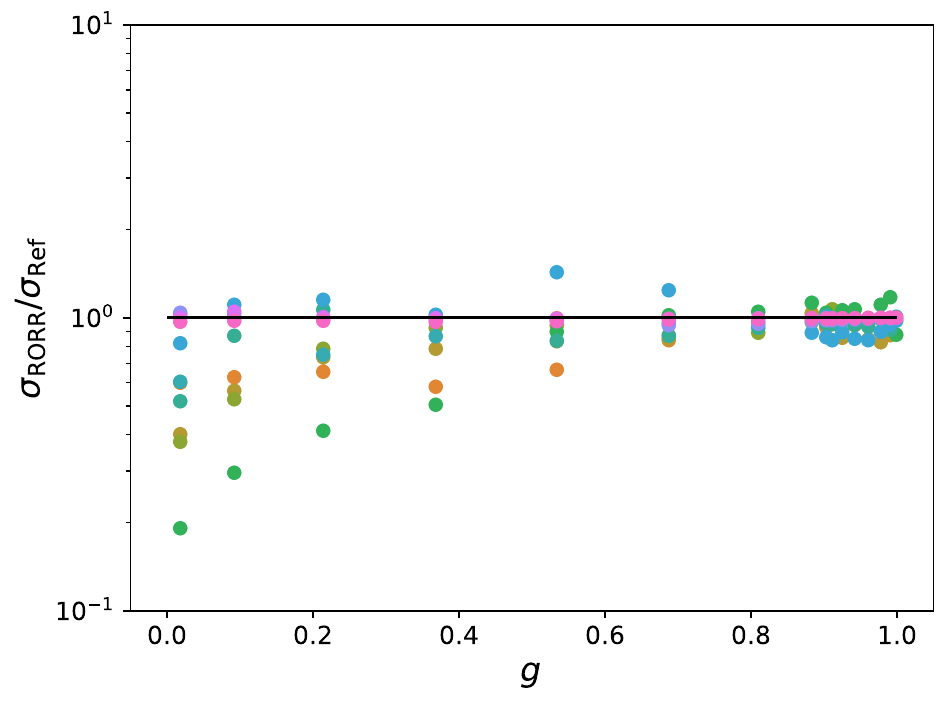}
    \includegraphics[width=0.48\linewidth]{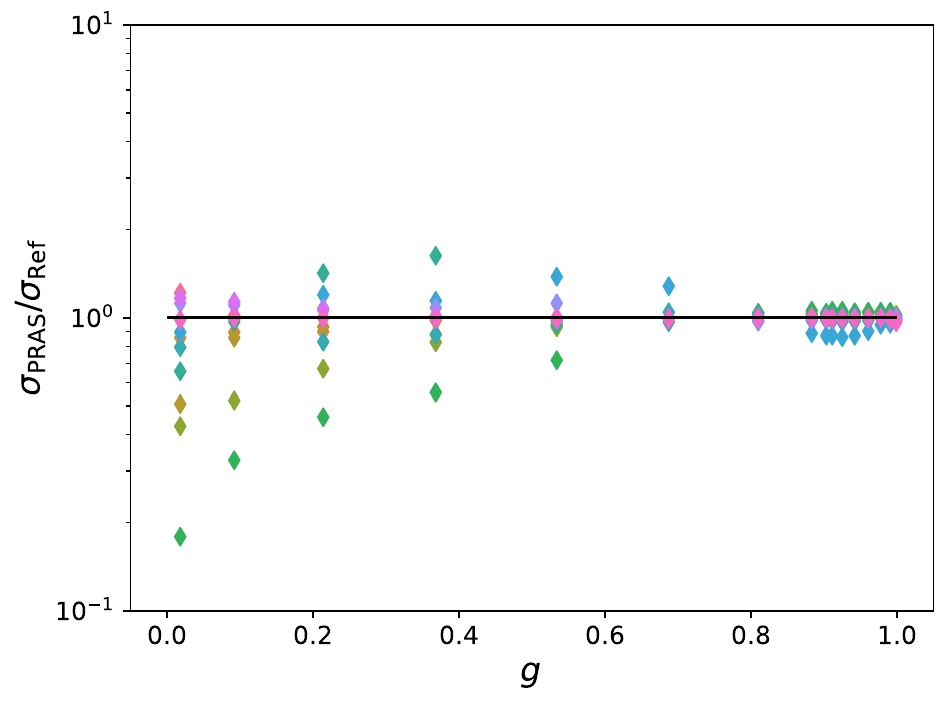}
    \caption{Comparison of different mixing methods to the reference pre-mixed (PM) values for the 11 wavelength bands  (colours) used in \citet{Kataria2013} for eight species at T = 1000 K, p = 1 bar (see text).
    Top left: Recovered cross sections at 8+8 Legendre quadrature nodes typically used in the petitRADTRANS retrieval model \citep{Molliere2019} for each method; PM (crosses), AEE (squares), RORR (dots), and PRAS (lines).
    Top right: Relative value between recovered AEE and PM cross sections.
    Bottom left: Relative value between recovered RORR and PM cross sections.
    Bottom right: Relative value between recovered PRAS and PM cross sections.}
    \label{fig:mix_ex}
\end{figure*}

\begin{figure*}
    \centering
    \includegraphics[width=0.48\linewidth]{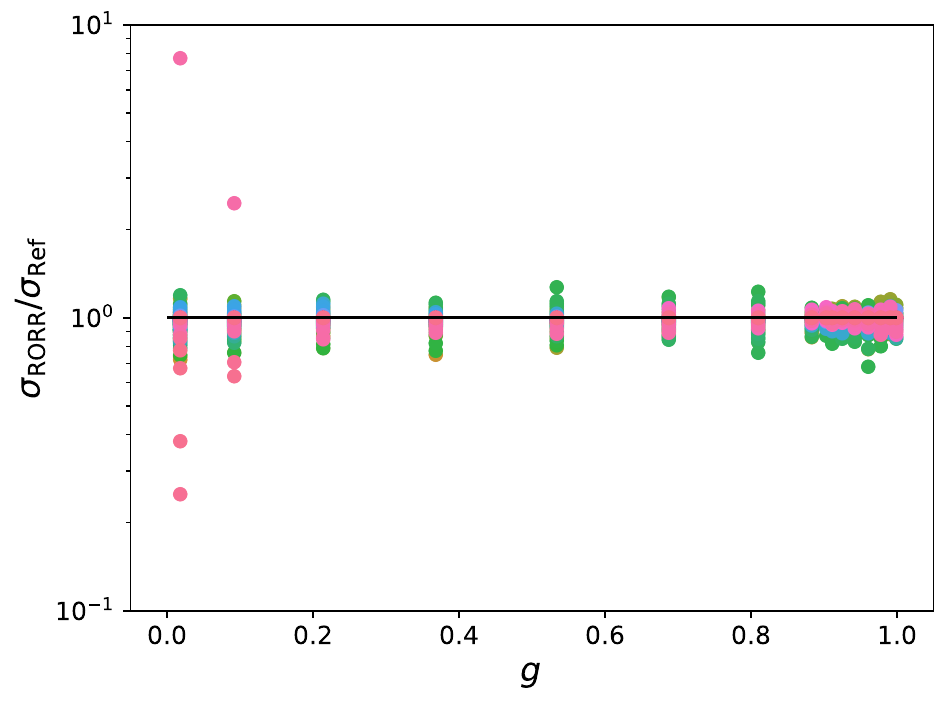}
    \includegraphics[width=0.48\linewidth]{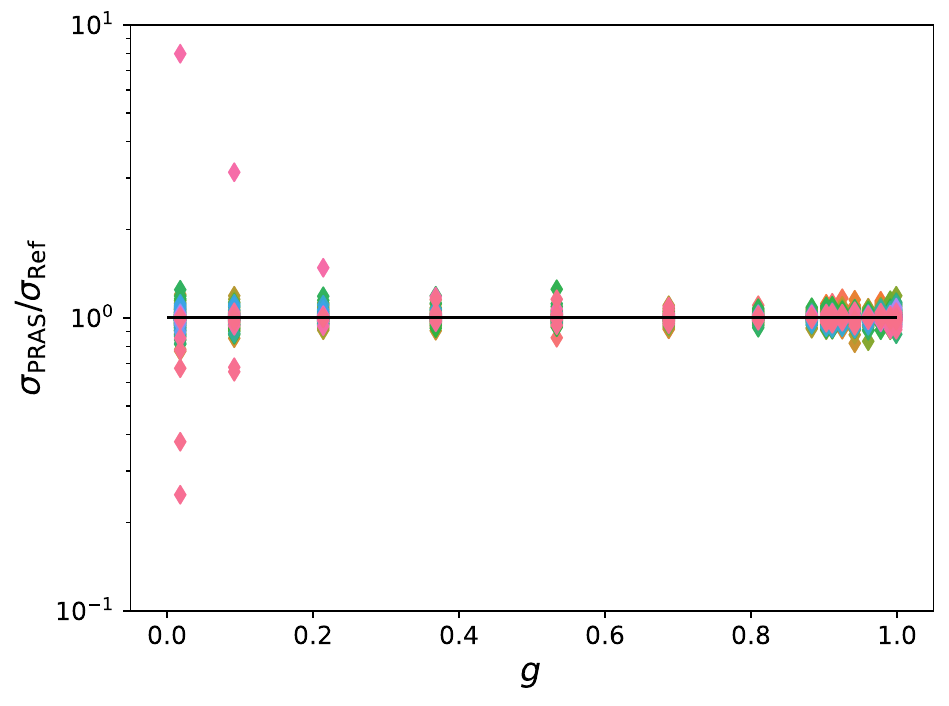}
    \caption{Comparison between the RORR and PRAS methods to the reference pre-mixed values for the $R$ = 100 spectral resolution wavelength bands (colours) for eight species at T = 1000 K, p = 1 bar (see text).
    I recover the 8+8 Legendre nodes typically used in the petitRADTRANS retrieval model \citep{Molliere2019}.
    Left: Relative value between recovered RORR and pre-mixed cross sections.
    Right: Relative value between recovered PRAS and pre-mixed cross sections.}
    \label{fig:mix_ex_R100}
\end{figure*}

As an addition to the methods outlined in the previous section, I propose a new, simple Monte Carlo probability density function (PDF) convolution method for producing randomly overlapped correlated-k opacities across a wavelength bin.
This method uses a polynomial reconstruction of the opacity CDF, which is then sampled to mix different species assuming random overlap between each species. 
I call this scheme the `Polynomial Reconstruction and Sampling (PRAS)' method, which is detailed in this section.

\subsection{Creating p-tables from opacity data}

The method starts with high fidelity cross-section tables, for example, such as those publicly available from DACE\footnote{\url{https://dace.unige.ch/opacityDatabase/}} database which contains cross-section data for several species at a resolution of $\Delta$$\nu$ = 0.01 cm$^{-1}$ or computed using various opacity calculators with input line lists \citep[e.g.][]{Grimm2015,Yurchenko2018,Agrawal2024}.
A specific wavelength range between a maximum and minimum wavelength is chosen to be the bin of interest. 
The number and size of each bin depends on the application, for example, gas giant exoplanet GCMs typically chose 11 \citep[e.g.][]{Kataria2013}, 30 \citep[e.g.][]{Showman2009} or 32 \citep[e.g.][]{Amundsen2014}.
Wavelength bands used for heating and cooling calculations must be chosen carefully, considering the trade-off of computational efficiency, accurately capturing opacity windows and the peak of the Planck function for the incident stellar and internal thermal radiation \citep[e.g.][]{Showman2009}.
Retrieval modelling can range between a spectral resolution of $\lambda/\Delta\lambda$ = $R$ = 100 and higher \citep[e.g.][]{Molliere2019}, which results in the number of wavelength bins in the 100s to 1000s.
For the correlated-k model to be accurate, the Planck function must be constant in the wavelength band.

The cross sections of gas species $a$ are sorted from lowest to highest value in this wavelength range, transforming the wavelength or wavenumber range into a $g$ $\in$ [0,1] space
\begin{equation}
    \sigma_{a}(g) = \sigma_{a}(\nu),
\end{equation}
where $\sigma_{a}(\nu)$ [cm$^{2}$ molecule$^{-1}$] are the cross sections in wavenumber, while $\sigma_{a}(g)$ [cm$^{2}$ molecule$^{-1}$] are the sorted cross sections in cumulative $g$ space.
This forms a representation of the normalised ($g$ $\in$ [0,1]), discretised inverse cumulative distribution function (CDF) of the opacity in the bin.
A polynomial (or spline) can then be used to fit an approximate functional form to the CDF of the logarithm of the cross sections
\begin{equation}
    \log_{10}\sigma_{a}(g) = \sum_{n=0}^{N_{p}} a_{n}P_{n}(g),
\end{equation}
where $N_{p}$ is the polynomial degree, $a_{n}$ the polynomial coefficients and $P_{n}$ the polynomial basis function.
Several basis functions can be used to fit the CDF, and in this work I use Chebyshev polynomials as an example to illustrate the method.
Alternatively, and in best practice, a global least squares (LSQ) B-spline or, following a similar idea to \citet{Saxner1984}, a Monotone Piecewise Cubic Hermite Interpolating Polynomial (PCHIP) can be fit to the CDF instead of a polynomial basis function. 
Chebyshev polynomials minimise oscillation error at the interval boundaries (Runge's phenomenon), while splines adapt better to local sharp gradients in opacity.
To try optimise the accuracy of spline fits, I split the knots into three $g$ sections, 0 $<$ $g$ $\leq$ 0.1, 0.11 $<$ $g$ $<$ 0.94 and 0.95 $\leq$ $g$ $<$ 1, with denser knot placements in the low and high $g$ regions. 
This follows similar strategies in the literature that attempt to better capture the potential strong gradients in the CDF near the extrema $g$ values \citep[e.g.][]{Showman2009, Kataria2013, Molliere2019}.

After the polynomial (or spline) fitting has been performed in a wavelength band, the nodes (or knots) and associated coefficients are stored for each band (p-tables), in a similar fashion to creating traditional correlated-k tables (k-tables), where the quadrature points, weights and k-coefficients are stored.
To enable interpolation in temperature and pressure (T-p) using the p-tables, the same node (or knot) positions must be used in a bin for each T-p opacity table required to be tabulated.
Coefficients for a given wavelength bin and species that are stored across a T–p grid can then be interpolated to reconstruct the polynomial or spline coefficients at a specific T-p at runtime.

Figure \ref{fig:cs_fit} shows a typical example of the p-table production method for a single \ce{H2O} wavelength band.
In this case, the Chebyshev polynomial fits the cross-section CDF across most of the $g$ value range to within 5\%, with larger inaccuracy at the extrema where the opacity gradients are steepest. 
There are some oscillatory structures due to the polynomial fit in the residuals.
The LSQ spline fitting is generally more accurate across the $g$ space, to within 2\%, except again at the extrema regions. 
The oscillatory structure in the spline fit is only obvious at the endpoints of the distribution where the gradient in cross section is steepest.
The piecewise, monotone properties of PCHIP method suggests it may be an excellent candidate for the PRAS scheme.
I find the residual error using the PCHIP method is sensitive to the number of interpolation points and selection of spline nodes (Figure \ref{fig:cs_fit}), but with optimal placement of spline nodes, less mean residual error than the LSQ method can be achieved.
I therefore use PCHIP as the primary method utilised throughout the paper.

\subsection{Random overlap using p-tables}

To randomly overlap two gases $a$ and $b$, weighted by their VMR, I randomly draw and store a number of samples, $n_{\rm s}$, from each of the polynomial functions for $a$ and $b$.
For example, for the $a$ component, for a random number $\zeta$ = [0,1], the sampled opacity, $\sigma_{\rm a, s}$ [cm$^{2}$ molecule$^{-1}$], is given by
\begin{equation}
    \log_{10}\sigma_{\rm a, s} = \sum_{n=0}^{N_{p}} a_{n}P_{n}(\zeta),
\end{equation}
which is repeated $n_{\rm s}$ times.
This sampling is also performed for gas $b$ using its own coefficients.
In the end, two arrays of size $n_{\rm s}$ are left containing the randomly sampled cross sections of gases $a$ and $b$.

From this random sampling of each polynomial, random overlap is achieved simply through summation of the sampled points weighted by the volume mixing ratio, $\chi$, of each species
\begin{equation}
    \sigma_{\rm mix} = \chi_{a}\sigma_{\rm s, a} + \chi_{b}\sigma_{\rm s, b},
\end{equation}
where $\sigma_{\rm mix}$ is the mixed cross sections and will be size $n_{\rm s}$.
This opacity mixture array is then sorted from minimum to maximum, which can then be used to interpolate the required k-coefficients for use in the correlated-k scheme, typically a chosen set of Legendre quadrature node values, for example, the 20 points typically used in NEMESIS \citep[e.g.][]{Irwin2008}, the 4+4 split method used in SPARC/MITgcm and PICASO \citep[e.g.][]{Showman2009, Mukherjee2023}, and the 8+8 split method used in petitRADTRANS \citep[e.g.][]{Molliere2019}.

This technique is easily generalised to an arbitrary number of gas species, $n_{\rm sp}$, through randomly sampling the polynomial of each gas species then summing 
\begin{equation}
    \sigma_{\rm mix} = \sum_{n=1}^{n_{\rm sp}}\chi_{n}\sigma_{\rm s, n}.
\end{equation}
In Appendix \ref{app:PRAS_lim}, I show this Monte Carlo sampling method is equivalent to convolution of the species' probability distribution functions. 

I can further increase accuracy and/or significantly reduce the required number of samples for the polynomial sampling through using Quasi-Monte Carlo (QMC) techniques.
I tested the scrambled Sobol and Halton sequences, as well as Latin Hypercube Sampling (LHS) \citep[e.g.][]{McKay1979}, which are quasi-random and sample a more uniformly spaced set of values between 0 and 1 than pure pseudo-random number generator sampling.
I found that utilising a scrambled LHS sequence enables a more accurate mixing of sampled polynomials than a pure pseudo-random number generator approach.
Using LHS enables a significant reduction in the required number of samples to retain good accuracy and random overlap properties. 
I found the Sobol and Halton sequences contained correlated artefacts and were less accurate.

\subsection{PRAS procedure summary}

I summarise below the practical steps to perform the PRAS scheme detailed above:
\begin{enumerate}
    \item Build and store p-tables of each species for specific wavelength bands and T-p points using the original high resolution cross-section data.
    \item During runtime, interpolate each p-table's polynomial or spline coefficients to the local T-p.
    \item Randomly sample each polynomial or spline function for each species a suitable number of times.
    \item Add each sample for all species together, then sort in ascending order.
    \item Determine the required $g$-values for the RT calculation and interpolate to find the required k-coefficients.
\end{enumerate}
Overall, this procedure follows a very similar structure to producing and mixing individual k-tables using the RORR method.

\section{Accuracy comparison}
\label{sec:acc}

\begin{figure}
    \centering
    \includegraphics[width=\linewidth]{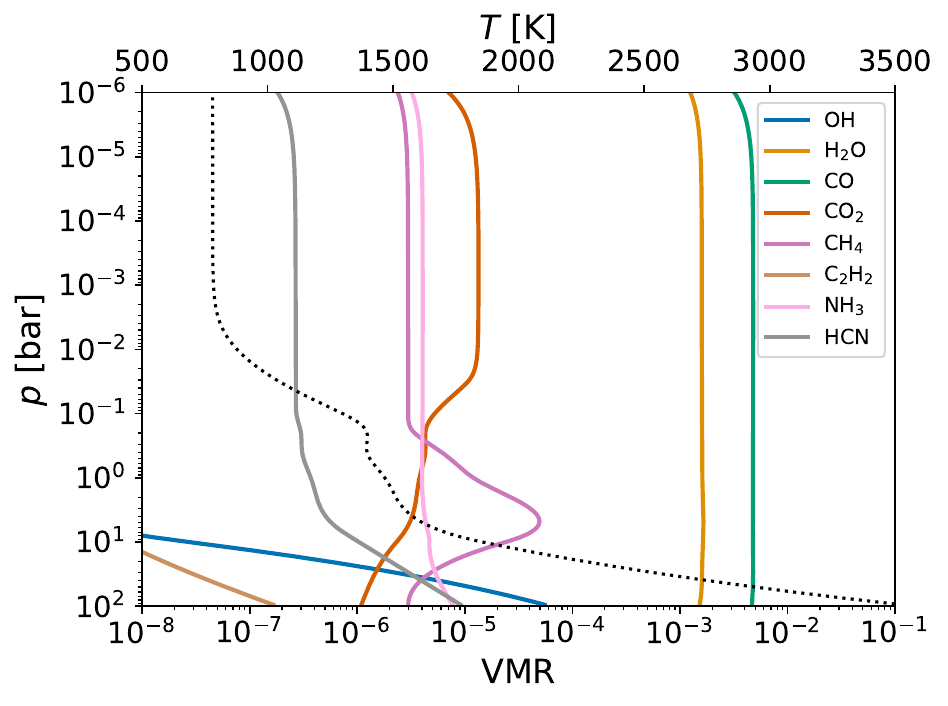}
    \caption{1D analytical temperature-pressure (T-p) for a globally averaged HD 189733b-like hot Jupiter atmosphere (grey dotted line) and VULCAN volume mixing ratio (VMR) of each species (coloured lines) considered in the emission spectra and heating rate comparison tests.}
    \label{fig:T_p_VMR}
\end{figure}

In this section, I compare the accuracy of the PRAS method to the PM, RORR and AEE methods for typical mixes of gases found in hot Jupiter atmospheres.
A useful aspect of PRAS is that it enables trading off between accuracy and speed through choices of polynomial degree or number of spline knots and number of random sample points.
The limiting factor for accuracy is how well the polynomial or spline approximates the cross-section CDF and the number of samples used to randomly overlap each polynomial.
I find the speed of computation is generally dominated by the number of samples used to randomly overlap, meaning that large degree polynomials or spline knots can be readily used to fit the cross-section CDF without too much computational overhead.
I therefore primarily explore how the number of samples affects the overall accuracy of the final k-coefficients produced by PRAS.
I note this is in contrast with the RORR method which requires an iterative species-by-species approach, and can only be tuned in accuracy and speed through changing the number and node selection of $g$ quadrature points.

\subsection{Individual cross section recovery}

I initially test the method through comparing the ability of each method to produce accurate mixed cross sections from calculated k-table or p-tables for individual species.
In the testing performed in \citet{Amundsen2017}, they find that RORR accuracy scales with number of nodes used, with 32 reproducing the reference LBL and RO results the best.
As a practical example, I consider eight of the main molecules used in the `mini-chem' kinetic chemistry model \citep{Tsai2022}, OH, \ce{H2O}, \ce{CO}, \ce{CO2}, \ce{CH4}, \ce{C2H2}, \ce{NH3} and HCN.

\subsubsection{Typical hot Jupiter GCM bands}

I take cross sections of each species from the DACE database at a temperature of 1000 K and pressure of 1 bar and generate k-tables and p-tables at the 11 GCM wavelength bands from \citet{Kataria2013} typically used in the SPARC/MITgcm.
This specific temperatures and pressures are equal to that of the cross-section database, avoiding any potential interpolation errors.
I use the FastChem chemical equilibrium code \citep{Stock2018} to generate representative VMRs of each species at $T$ = 1000 K, $p$ = 1 bar with 10$\times$ solar elemental abundances ([M/H] = 1), resulting in VMRs for OH: 2.7 $\cdot$ 10$^{-14}$, \ce{H2O}: 4.8 $\cdot$ 10$^{-3}$, \ce{CO}: 8.3 $\cdot$ 10$^{-4}$, \ce{CO2}: 6.9 $\cdot$ 10$^{-6}$, \ce{CH4}: 4.1 $\cdot$ 10$^{-3}$, \ce{C2H2}: 4.0 $\cdot$ 10$^{-12}$, \ce{NH3}: 1.1 $\cdot$ 10$^{-5}$ and HCN: 6.6 $\cdot$ 10$^{-9}$.
For the accuracy comparison, I generate a PM k-table using the cross-section data, weighted by the VMR of each species which I use as the benchmark values.
I then mix individual species k-tables using RORR and AEE with 8+8 Legendre nodes, typically used in the petitRADTRANS retrieval model \citep{Molliere2019}, as well as PRAS with 32 spline knots and 10,000 samples.
The relative difference between the reference PM values and individual mixed values are then calculated for each Legendre node.
In an RT model context, the cross section associated with each Legendre node would then be used in the RT calculation, making the accuracy of recovering cross sections at these points an important consideration when developing a mixing method.

In Figure \ref{fig:mix_ex}, I show the results for each method in each wavelength band and the relative fraction between the mixed results and the reference PM values.
Overall, the PRAS method shows similar k-coefficient recovery characteristics to the RORR method, with similar structures in the relative fraction between the mixed values and the reference values.
Most bands using PRAS show an overall tighter spread in accuracy compared to RORR, especially in the medium to low $g$ value regions ($g$ $<$ 0.9).
The AEE method shows differences increasing at the lower values of $g$, where the grey assumption has the most impact, however, the relative values remain reasonable at larger values of $g$.
This is not unexpected, as in this test, I have chosen both RORR and PRAS parameters that produce high accuracy with respect to the random overlap assumption.
The differences to the reference values here therefore represent errors associated with the random overlap assumption itself. 
However, this test shows that the PRAS method produces similar results, with comparable errors to RORR, due to the random overlap assumption, compared to the PM reference values.

In Appendix \ref{app:mix_ex}, I present the recovery results for more extreme T-p points available from the original cross-section data.
In these examples, the PRAS method also shows very similar error characteristics to RORR, with it producing similar structures for each mixed k-coefficient.
The PRAS method generally produces a tighter margin of error across $g$ space compared to RORR, especially in the lower temperature and pressure case ($T$ = 50 K, $p$ = 10$^{-8}$ bar), suggesting PRAS is better capturing the mixed opacity distribution in this regime.
In the $T$ = 50 K, $p$ = 1000 bar case, both methods struggle to reasonably reproduce the pre-mixed results, suggesting higher correlation of lines in these conditions, likely due to pressure broadening of each line.
However, such an extreme T-p point (very cold at high pressure) is rarely encountered in a real atmosphere.

\subsubsection{Typical retrieval bands}

I now repeat the above exercise but with a wavelength range of 0.2-30 $\mu$m at a spectral resolution of $R$ = 100, a typical lower end resolution used in contemporary retrieval modelling for substellar atmospheres.
I recover cross sections at the 8+8 split Legendre nodes typically used in the petitRADTRANS retrieval model \citep{Molliere2019}.
I produce k-tables and p-tables for all species at a temperature of 1000 K and pressure of 1 bar.
For the p-tables I use 32 spline knots.

In Figure \ref{fig:mix_ex_R100} I show the results of this test, comparing the RORR and PRAS methods to reference PM values.
Similar to the previous test, the PRAS method shows consistent error characteristics to the RORR method.
Overall, PRAS is slightly tighter towards the PM reference line than the RORR method in the mid and high $g$ value range in most bands.
This suggests PRAS is able to recover slightly more accurate k-coefficients with respect to the reference pre-mixed cross sections than RORR.

\subsection{Emission spectra accuracy}

\begin{figure*}
    \centering
    \includegraphics[width=0.49\linewidth]{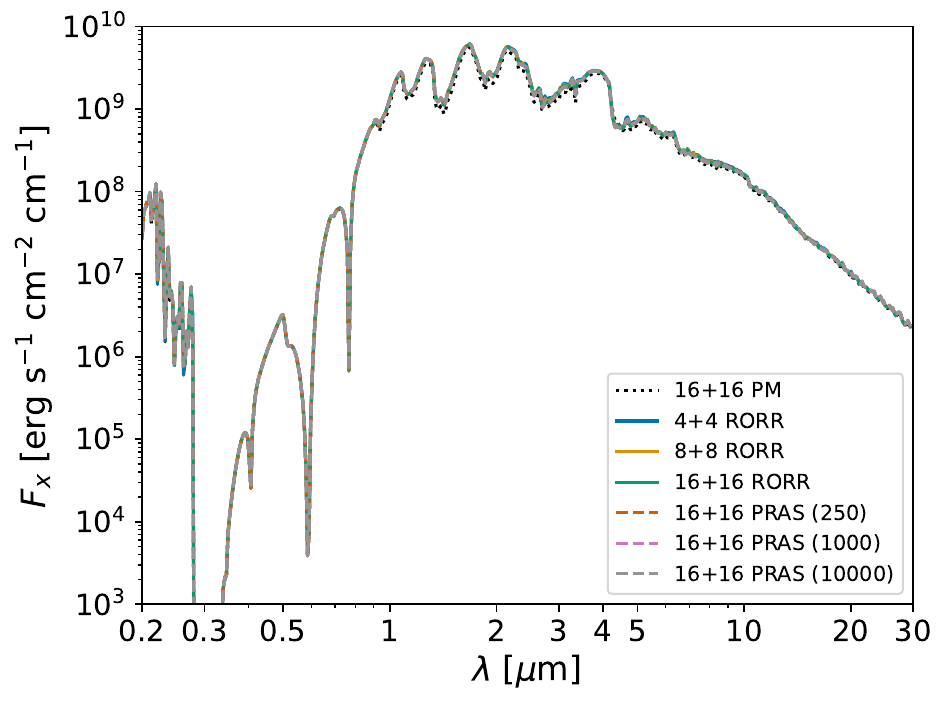}
    \includegraphics[width=0.49\linewidth]{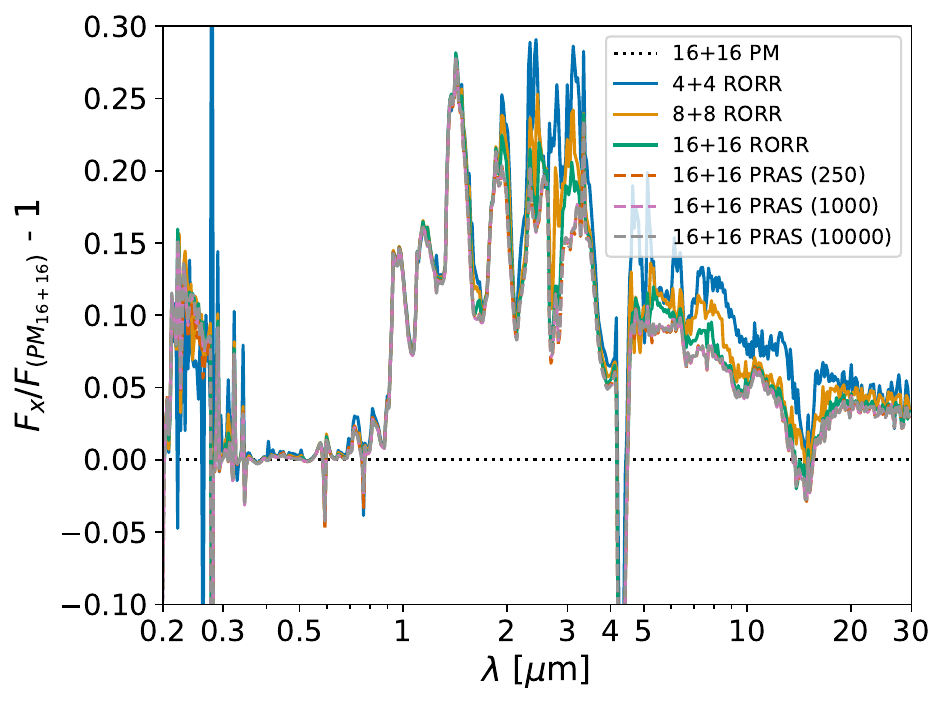}
    \caption{Comparison between different mixing methods for the emission spectra test for a spectral resolution of $R$ = 100.
    The number in the brackets for the PRAS label denotes the number of random samples used to mix the opacity.
    Left: Planetary emission flux of each method.
    Right: Relative value compared to the 16+16 PM method reference spectra.}
    \label{fig:olr}
\end{figure*}

In this section, I test the accuracy of the outgoing planetary emission spectra fluxes of each mixing method, through calculating an emission spectra of a given one-dimensional T-p and VMR profile.
I use the analytical picket-fence T-p profile scheme from \citet{Parmentier2014} and \citet{Parmentier2015}, with input planetary parameters of the hot Jupiter HD 189733b taken from \citet{Addison2019}, assuming a globally averaged T-p profile with full energy redistribution ($f$ = 1/4).
I then use the 1D photochemical model VULCAN \citep{Tsai2021}, with the default C-H-O-N chemical network at 10x solar metallicity atmosphere, and a constant $K_{\rm zz}$ = 10$^{8}$ cm$^{2}$ s$^{-1}$.
I model 200 layers between an upper pressure of 100 bar and lower pressure of 10$^{-8}$ bar.
For clarity and convenience, photochemical reactions are excluded.
I create temperature and pressure dependent k-tables and p-tables at a spectral resolution of $R$ = 100 between 0.2-30 $\mu$m (503 total bands) for all considered species, with the p-tables using 32 spline knot points each.
I mix the opacity of ten species OH, \ce{H2O}, \ce{CO}, \ce{CO2}, \ce{CH4}, \ce{C2H2}, \ce{NH3}, HCN, Na and K.
I assume Na and K are constant and quenched from the deep layers at 10x their solar VMR values at Na: 10$^{-5}$ and K: 10$^{-6}$ \citep[e.g.][]{Woitke2018}.
Figure \ref{fig:T_p_VMR} presents the vertical T-p and VMR structure of my test atmosphere.

I produce a reference emission spectra using a PM k-table at the T-p and VMRs of the vertical structure, and make several different comparisons between various methods:
\begin{itemize}
    \item 4+4, 8+8 and 16+16 distributed Legendre node k-tables with RORR.
    \item PRAS method recovering 16+16 Legendre nodes with a low (250), medium (1000) and high (10,000) number of samples.
\end{itemize}
This aims to test the potential differences seen between RORR and PRAS when calculating emission spectra as well as how well it compares to the reference PM spectra.
This PM spectra may contain interpolation errors stemming from interpolating in T-p from the original DACE database cross-section data as seen in \citet{Amundsen2017}, but still serves as useful reference for each of the overlap methods.

In the RORR and PRAS cases, I use a split Legendre node scheme commonly used in the literature, where 4+4 corresponds to a split at $g$ = 0.95 \citep{Showman2009, Mukherjee2023}, 8+8 with a split at $g$ = 0.9 \citep{Molliere2019} and a custom 16+16 with a split at $g$ = 0.9.
These schemes attempt to better capture contribution at high $g$ values of the cross-section CDF where the opacity values and gradient are largest.

Figure \ref{fig:olr} presents the results of the emission spectrum test.
All emission spectra are highly similar, with no easily discernible difference between each method by eye.
However, from the relative values, in the RORR cases, it is clear that increasing the number of quadrature nodes generally leads to less errors, with increasing numbers of node shifting closer towards the reference PM fluxes.
In the PRAS case, the number of samples does not substantially alter the end result, with all three values showing highly similar relative values to the 16+16 RORR result.
The PRAS method is also generally more accurate than the 8+8 and 4+4 RORR results even with the lowest number of samples considered (250), with the relative value only differing by around 2$\%$ from the 16+16 RORR results.
However, I find, in this test case, the random overlap assumption present in PRAS and RORR can underestimate the net opacity in most bands in the near- and mid- infrared, leading to a departure of up to $\sim$25\% from the reference PM results.
Despite this, I find PRAS typically performs the best of the overlap methods, with the least amount of error of all schemes.
Overall, this result suggests that PRAS is particularly suitable for producing emission spectra, especially when increased accuracy is required, such as when performing retrieval models or post-processing output from 1D or 3D models.

\subsection{Flux and Heating rates}

\begin{figure*}
    \centering
    \includegraphics[width=0.49\linewidth]{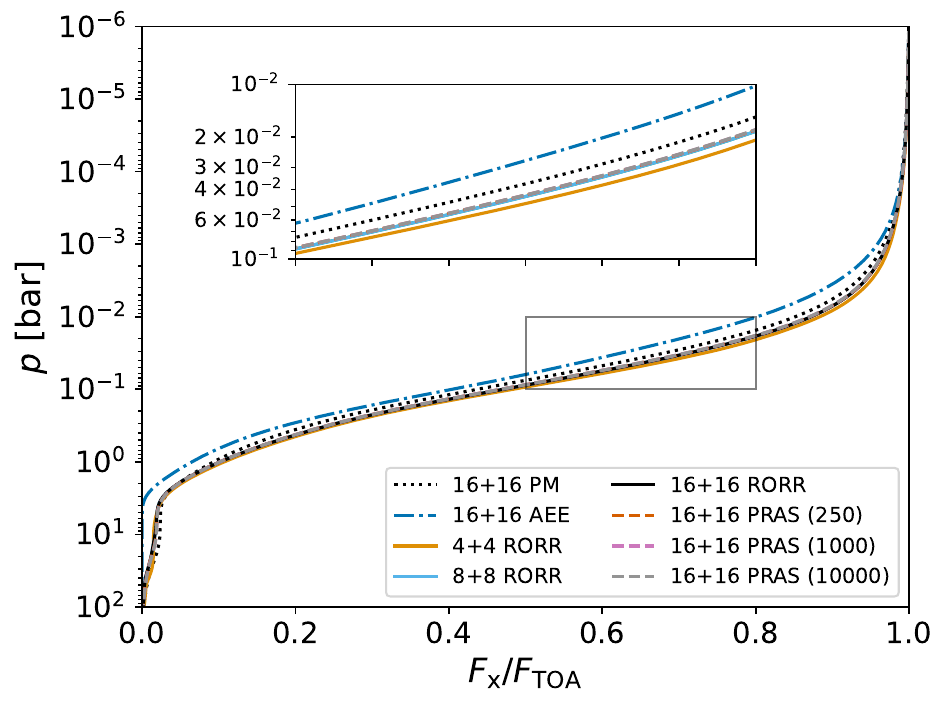}
    \includegraphics[width=0.49\linewidth]{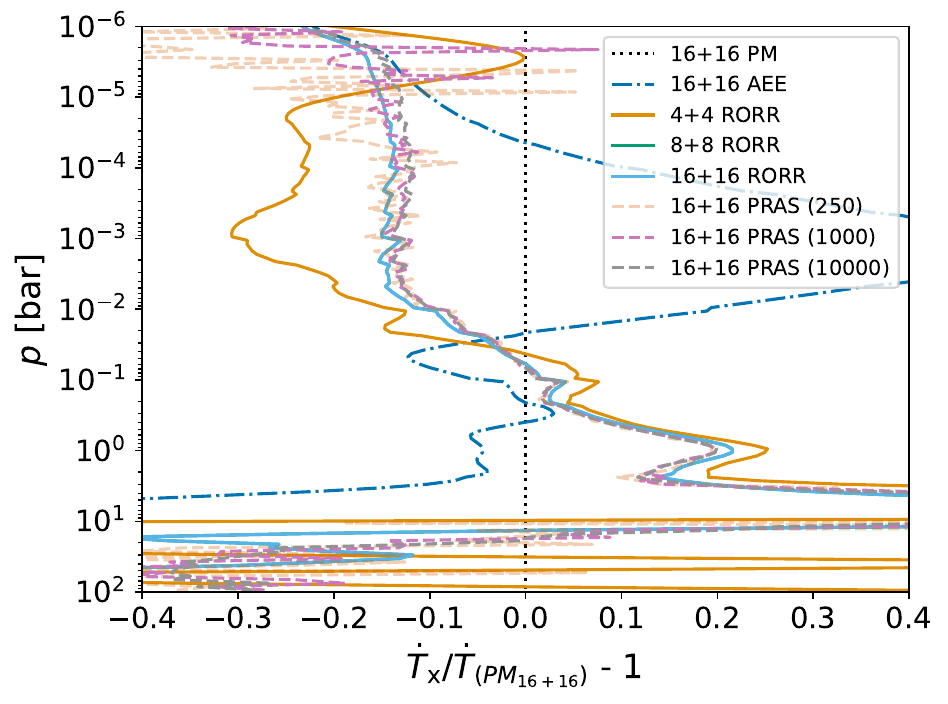}
    \includegraphics[width=0.49\linewidth]{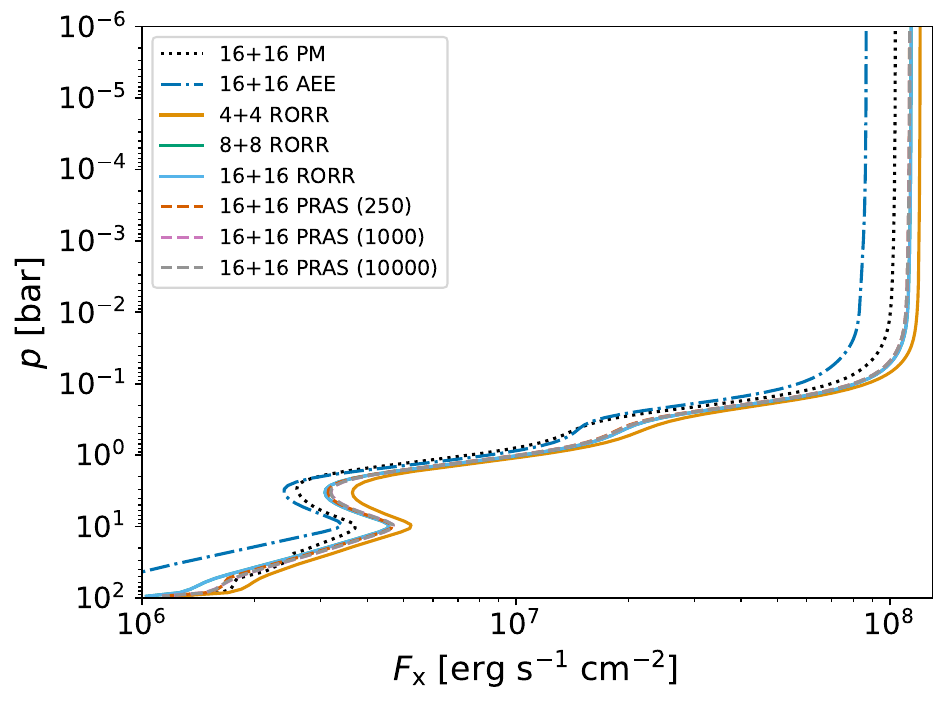}
    \includegraphics[width=0.49\linewidth]{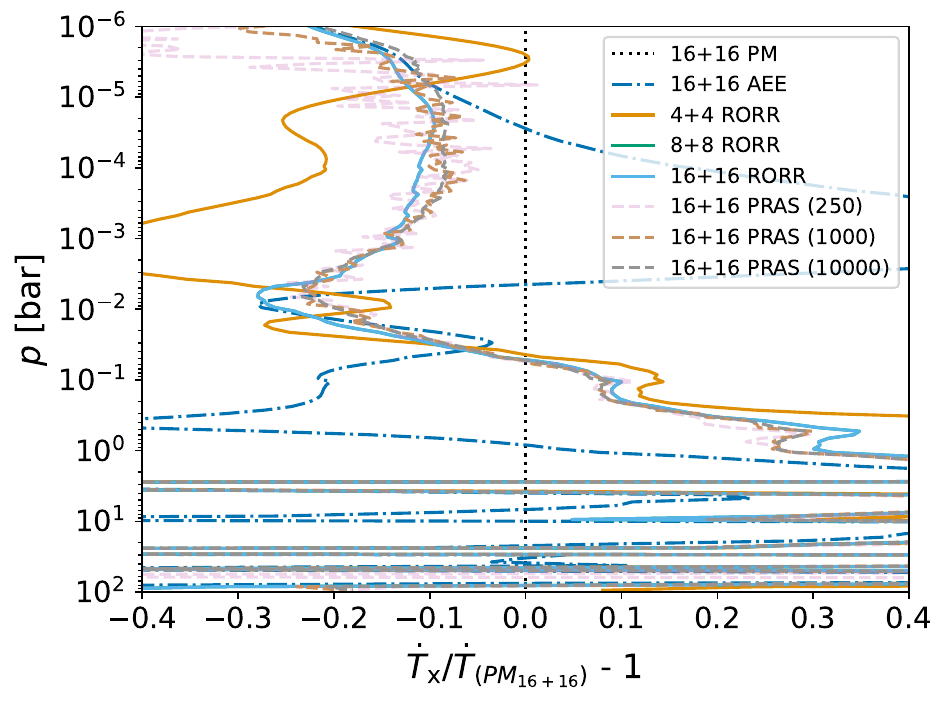}
    \caption{Flux (left) and heating rate (right) calculation tests for the shortwave (top row) and longwave (bottom row) schemes.
    The number in the brackets for the PRAS label denotes the number of random samples used to mix the opacity.
    Overall, PRAS is able to recover vertical fluxes highly comparable to the 16+16 RORR method using 1000 or 10,000 samples.
    For the heating rates, the error characteristics from the number of samples used in the PRAS method shows similar rates are recovered with less samples, but with an overall increased variance.
    However, all PRAS calculations show similar or more accurate overall heating rate errors compared to the PM reference than the 4+4 and 8+8 RORR and AEE schemes.}
    \label{fig:flux_heat}
\end{figure*}

For the internal atmospheric fluxes and heating rate comparison, I use the 32 wavelength bins from \citet{Amundsen2014}, with the same T-p and VMR profile from the emission spectra case.
I assume a constant atmospheric heat capacity of $c_{\rm p}$ = 1.14$\cdot$10$^{8}$ erg g$^{-1}$ K$^{-1}$, typical of a 10x solar, \ce{H2} and He dominated atmosphere.
I make six different comparisons to the reference PM results with the following methods
\begin{itemize}
    \item 16+16 Legendre nodes with AEE.
    \item 4+4, 8+8 and 16+16 Legendre nodes with RORR.
    \item PRAS recovering 16+16 Legendre nodes with a low (250), medium (1000) and high (10,000) number of samples.
\end{itemize}
In this way, I can compare the accuracy differences between AEE and PRAS when producing heating rates in the atmosphere for typical hot Jupiter GCM conditions, as well as test how accuracy is affected from the number of samples assumed in PRAS.

To calculate incident fluxes entering the atmosphere from an irradiating source such as a star (also known as shortwave radiation in planetary science contexts), I use a simple 1D direct beam only calculation at each altitude level $z$
\begin{equation}
    F_{\rm sw}(z) = \sum_{b=1}^{n_{b}}\mu_{*}F_{\rm TOA}(b)\sum_{g=1}^{n_{g}}\exp\left[-\tau(z,b,g)/\mu_{*}\right]w_{g},
\end{equation}
where $\mu_{*}$ is the cosine zenith angle (here 1/$\sqrt{3}$), $\tau$($z,b,g$) is the optical depth at each altitude level $z$, for band $b$ and $g$ node and $w_{g}$ the weight of the $g$ quadrature node. 
To compare each method fairly, I normalise the shortwave fluxes by the top of atmosphere flux.
For the internal, thermal flux calculation (also known as longwave radiation in planetary science contexts), I use the 1D short characteristics with linear interpolants method \citep{Olson1987} with eight streams following the \citet{Hogan2024} Gauss–Laguerre emission angles and weights.
I ignore any scattering effects in the shortwave or longwave flux calculation.

The heating rates, dT/dt [K s$^{-1}$], from the shortwave and longwave radiation are given as \citep[e.g.][]{Amundsen2014, Malik2019}
\begin{equation}
    \frac{dT_{\rm sw}}{dt} = \frac{g_{\rm surf}}{c_{\rm p}} \frac{\Delta F_{\rm sw}}{\Delta p},
\end{equation}
and
\begin{equation}
 \frac{dT_{\rm lw}}{dt} = \frac{g_{\rm surf}}{c_{\rm p}} \frac{\Delta F_{\rm lw}}{\Delta p},
\end{equation}
respectively, where  $g_{\rm surf}$ [cm s$^{-2}$] is the surface gravity of the planet.
I use the heating rates results from the 16+16 PM calculations as the reference values, and calculate the relative values for each method to this reference flux.

Figure \ref{fig:flux_heat} shows both the shortwave and longwave vertical fluxes and heating rates produced by each method.
For the shortwave fluxes, my results show that the 4+4 and 8+8 RORR results gradually converge towards the 16+16 RORR values, suggesting an increase in accuracy with number of quadrature points used in the RORR method.
These RORR results show a general trend of underestimating the opacity as less quadrature points are used, suggesting that the larger opacity regions of the opacity CDF, higher $g$, are less well represented with a smaller number of $g$ values.
For the PRAS results, each number of samples produces highly similar results, and show similar characteristics to the RORR results.
The AEE results generally overestimate the opacity of the atmosphere, producing significant differences, $>$10\%, compared to the reference PM result.

For the shortwave heating rate relative values results, again a similar trend is seen in the RORR results, with the heating rates converging to the 16+16 results as more quadrature points are used.
The PRAS results are typically similar to the 16+16 RORR results to within $\sim$5\%, but are shifted more towards the reference PM values.
Notably, PRAS is closest to the reference in the very upper atmosphere $p$ < 10$^{-2}$ bar and deep in the atmosphere where gradients in the \ce{CH4} abundance are present (Figure \ref{fig:T_p_VMR}).
I find accuracy scales with the number of samples, with 250 producing a more noisy heating rate structure, especially in the upper atmosphere, while 1000 and 10,000 samples produce similar overall structures to the 250 samples but with less noisy structures.
This suggests that PRAS can be tuned as a function of pressure, with less samples for higher pressure regions and more for lower pressure regions.

For the longwave fluxes, a similar pattern is seen to the shortwave fluxes.
The RORR results again converge towards the 16+16 RORR values, suggesting adding more quadrature points increases accuracy.
PRAS again shows similar flux values for all number of samples, and is within $\sim$5\% to the 16+16 RORR results.
Again, due to the random overlap assumption, the opacity is generally underestimated, leading to an overall larger net flux with pressure in the atmosphere compared to the reference PM simulation.
The net OLR is overestimated in the RORR cases, by $\lesssim$10\%, in line with the previous sections spectral results.
Interestingly, at pressures of $\sim$2 bar, the AEE and PM fluxes agree well.
This is where \ce{CH4} becomes more abundant in the atmosphere (Figure \ref{fig:T_p_VMR}), suggesting a stronger dominance of \ce{CH4} on the net opacity compared to other molecules.

For the longwave tests, the PRAS results show a similar heating rate structure with number samples, with the variance in the results increasing with less samples.
This is particularly evident in the upper atmosphere ($p$ < 10$^{-2}$ bar), where 250 samples is significantly more noisy compared to the 1000 and 10,000 sample run.
Overall, I find PRAS most beneficial in the upper atmosphere ($p$ < 10$^{-2}$ bar), where it is $\sim$5\% more accurate that the 16+16 RORR method at reproducing the reference PM results.

\section{Computational efficiency}
\label{sec:speed}

\begin{table}[]
    \centering
    \caption{Computational speed test of the AEE, RORR and PRAS methods, repeating the mixing of opacities in the OLR comparison case 10 times, for 100,800 total mixing calculations each.
    The value in brackets for the PRAS method is the number of random samples used in the overlap.}
    \begin{tabular}{c|c|c} \hline \hline
     Method & Total time (s) & Relative time \\ \hline
     16+16 AEE & 78 & 1.0 \\
     4+4 RORR & 85 & 1.1 \\
     8+8 RORR & 116 &  1.5 \\
     16+16 RORR & 231 & 3.0 \\
     16+16 PRAS (10) & 47 & 0.6 \\
     16+16 PRAS (250) &  106 & 1.4 \\
     16+16 PRAS (1000) & 286 & 3.7 \\
     16+16 PRAS (10,000) & 2664 & 34.1 \\
     \hline \hline
    \end{tabular}
    \label{tab:speed}
\end{table}

In this section, I provide a basic analysis of the speed comparison between the AEE, RORR and PRAS methods.
The advantage of the PRAS method is that the computational cost of mixing species remains relatively constant with number of $g$ nodes required for the RT calculation.
This is due to the simple calculations (sampling) contained in the PRAS method and its representation of the full CDF structure using polynomial or spline coefficients.
In contrast, the RORR method quickly becomes computationally expensive the more $g$ nodes are required, due to the need for considering the weighted convolution of each $g$ node, especially in contexts where mixing is frequently required for a larger number of molecules and atoms such as in retrieval modelling.

In addition, the PRAS method is an example of an `embarrassingly parallel' problem, as the sampling of the polynomial functions for each species may be carried out independently in parallel and summed together at the end of the parallel computation.
Due to this property, the gain from parallelisation is likely to be near linear with processor count and amenable to GPU acceleration.
This would enable a larger numbers of samples to be used to mix the p-tables to reach the required accuracy, without sacrificing runtime.

To set up a computational speed comparison, I used the same setup as the OLR accuracy test at $R$ = 100 with 503 wavelength bands.
I then use the AEE and RORR methods for the eight species considered above for three different Legendre node numbers, split 4+4 \citep{Showman2009, Mukherjee2023}, split 8+8 \citep{Molliere2019} and 16+16 as used in the literature.
For the PRAS method, I use 32 spline knots with various numbers of samples  during the mixing stage and recover 16+16 Legendre node values.
In testing, I found that the computational efficiency of the PRAS method scales more rapidly with number of random samples, rather than the number of polynomial coefficients or spline knots.
This is due to the end step requiring to sort the convolved random samples, and with larger numbers of samples this sorting step can be a significant portion of the computational burden.
This suggests that the evaluation of the functional representation of the opacity CDF is not the main computational expense in the PRAS methods, but rather the overall number of samples (e.g. for the same number of samples, a change from 32 to 16 spline knots does not reduce the computational burden by half.).

I force mixing of each species in each layer and band to be performed 10 cycles and time the total execution time\footnote{Timings were performed on an Apple M4 Mac mini (2024)}.
I attempt as many numerical optimisations as possible in the python code for each method, utilising optimised libraries when available.
In Table \ref{tab:speed}, I present the total computational time and relative time to the AEE method.
Overall, I find that PRAS is comparable in speed to the RORR method, except when a large number of samples (e.g. 10,000) is used.
The number of samples used to perform the Monte Carlo integration is the main computational limiting factor.
However, fully optimised versions or the potential impact of parallelisation of each scheme cannot be guaranteed, therefore the values in Table \ref{tab:speed} should be taken as indicative rather than absolute performance.

\section{Discussion}
\label{sec:disc}

In my computational speed test, I only provided a basic analysis using a python implementation used to test the overall accuracy of the PRAS methodology, rather than optimised for computational efficiency.
Conclusions regarding speed comparisons between each method may therefore differ from this study when implemented in compiled languages using optimised numerical libraries.
My initial tests show that the PRAS method is similar in computational burden to the RORR method.
\citet{Amundsen2017} suggest that this may not be fast enough for implementation inside a GCM simulation.
However, due to the increased accuracy provided by PRAS, especially for outgoing thermal flux spectra, PRAS may be useful for retrieval, one-dimensional atmospheric modelling and multi-dimensional simulation post-processing efforts.
These models may take full advantage of the ability of PRAS to recover a larger number of mixed k-coefficients than RORR at a comparable speed. 

I suggest several improvements to the current basic scheme used here can be made with careful considerations
\begin{itemize}
    \item Optimal or automatic selection of spline knots to further increase the accuracy of the cross-section CDF representation in a given band.
    \item Techniques to increase accuracy of the polynomial or spline fit near the endpoints of the $g$ space, especially for the $g$ $<$ 0.1 and $g$ $>$ 0.9 regions.
    \item Using sampling biasing methods to reduce variance in the Monte Carlo sampling stage, however, a correct weighting scheme for each sample would need to be used. 
    \item Techniques to sample the spline functions with fewer total samples while retaining good accuracy, such as temperature and pressure dependent sampling numbers.
\end{itemize}
Overall, there is a significant scope to increase both the accuracy and efficiency of the PRAS scheme.
Another promising avenue to explore are machine learning techniques that could accelerate the recovery of mixed cross sections \citep[e.g.][]{Schneider2024}.

In the pure random sampling method, the root mean squared error (RMSE) in the convolution scales with number of samples, following standard Monte Carlo convergence rates
\begin{equation}
    {\rm RMSE} \propto \frac{1}{\sqrt{N}}.
\end{equation}
In this study, I attempted to use Latin Hypercube Sampling (LHS) to more evenly spread sample values across $g$ $\in$ [0,1] without violating random overlap.
I find this technique significantly reduces the required number of samples to accurately perform the Monte Carlo convolution.
These stratified sampling schemes often have a faster RMSE convergence rate
\begin{equation}
     {\rm RMSE} \propto \frac{1}{N^{q}},
\end{equation}
where in some idealised cases $q$ $\geq$ 1 gives a linear or better decrease in error with sample number.
However, care must be taken when using stratified sampling techniques to avoid correlated artefacts in the final result.

In this study, I typically tested 250, 1000 and 10,000 samples, assuming a linear RMSE dependence ($q$ = 1) from the LHS approach, this results in around a 1\%, 0.1\% and 0.01\% RMSE from the Monte Carlo sampling alone.
Assuming a range of 1-10\% error in the CDF spline fitting, I can expect around a total 1-10\% error with 250 samples compared to the idealised perfect case.
\citet{Garland2019} suggest that the error associated with the correlated-k assumption is also around 1-5\% in an exoplanet atmosphere context compared to LBL calculations.
Overall, this suggests that the total error associated with the PRAS random overlap mixing along with the correlated-k method is very low, in the $\leq$ 10\% range.
However, I note that while PRAS is well suited for random overlap, increasing accuracy when partial or correlated opacities are dominant will require modified sampling strategies or hybrid approaches, which I leave for future work.

With the ability to more easily mix larger numbers of Gaussian quadrature nodes than RORR, PRAS may be important in retrieval contexts, especially in the era of high precision brown dwarf emission spectra provided by JWST \citep[e.g.][]{Miles2023, Biller2024, McCarthy2025, Matthews2025}.
As seen in my emission spectra test, I suggest PRAS is a more accurate representation of the opacity structure than RORR, and is able to more easily mix and recover more quadrature points than the commonly used 4+4 and 8+8 quadrature schemes.
In a retrieval context, a 2\% change in the accuracy of the opacities can translate to a significant shift in the overall spectra produced by forward models \citep[e.g.][]{Baudino2017, Leconte2021}.
Recent studies have shown the sensitivity of the opacity scheme used in different retrieval models to the end result \citep[e.g.][]{Garland2019, Barstow2020}, as well as accounting for general uncertainty between models \citep[e.g.][]{Nixon2024} of which their opacity schemes are a part.

\section{Conclusions}
\label{sec:conc}

I propose a simple and intuitive Monte Carlo-based method for mixing the opacity of species in a wavelength band assuming random overlap.
Due to the core of the method using polynomial or spline fitting and sampling I call this method polynomial (or spline) reconstruction and sampling (PRAS).
In the limiting case where the opacity cumulative distribution functions (CDFs) of all species are perfectly represented by polynomials or splines (i.e. zero residuals), and infinite random sampling is performed, the PRAS method asymptotically achieves exact modelling of random overlap.
In this limit, the method converges to the exact convolved opacity distribution with no additional mixing error.

PRAS is inherently tunable in accuracy and computational speed from the goodness of fit for the polynomial or spline fit to the opacity CDF and the total number of random samples of the polynomial or spline fits.
Overall, due to the similar computational efficiency to RORR, I suggest the current PRAS method presented in this study is most suited for 1D modelling, retrieval and post-processing applications when greater accuracy is required for the RT calculations than RORR.
Due to the scalability of PRAS, recovering larger numbers of quadrature nodes is simple, for example the 16+16 nodes used in this study, enabling an increase in overall accuracy of the opacity representation in the atmosphere for RT calculations.

My results suggest that PRAS is around 2-5\% more accurate than RORR at reproducing reference PM vertical flux and heating rates, with both methods using the same 16+16 quadrature nodes.
I find PRAS recovering 16+16 nodes is generally more accurate than the typically used 4+4 and 8+8 node RORR methods in the substellar modelling literature.

My basic computational speed tests suggest the PRAS method may be too slow for time dependent GCM applications, however, I suggest there is significant future scope to tune and improve the PRAS method in both accuracy and speed.
Should this be achievable in the future, increased accuracy beyond the AEE method may be feasible for large scale substellar models that require calculation of heating rates such as GCMs.
Several Python codes outlining the PRAS methodology and reproduction of the tests performed in this study are available on GitHub\footnote{\url{https://github.com/ELeeAstro/PRAS}}.

\begin{acknowledgements}
E.K.H. Lee is supported by the CSH through the Bernoulli Fellowship.

\end{acknowledgements}

\bibliographystyle{aa}
\bibliography{bib.bib}

@ARTICLE{Stein_2024,
       author = {{Stein}, Robert F. and {Nordlund}, {\r{A}}ke and {Collet}, Remo and {Trampedach}, Regner},
        title = "{The Stagger Code for Accurate and Efficient, Radiation-coupled Magnetohydrodynamic Simulations}",
      journal = {\apj},
         year = 2024,
        month = jul,
       volume = {970},
       number = {1},
          eid = {24},
        pages = {24},
          doi = {10.3847/1538-4357/ad4706},
archivePrefix = {arXiv},
       eprint = {2405.02483},
 primaryClass = {astro-ph.IM},
       adsurl = {https://ui.adsabs.harvard.edu/abs/2024ApJ...970...24S}
}

@ARTICLE{Vogler_2005,
       author = {{V{\"o}gler}, A. and {Shelyag}, S. and {Sch{\"u}ssler}, M. and {Cattaneo}, F. and {Emonet}, T. and {Linde}, T.},
        title = "{Simulations of magneto-convection in the solar photosphere.  Equations, methods, and results of the MURaM code}",
      journal = {\aap},
         year = 2005,
        month = jan,
       volume = {429},
        pages = {335-351},
          doi = {10.1051/0004-6361:20041507},
       adsurl = {https://ui.adsabs.harvard.edu/abs/2005A&A...429..335V}
}

@ARTICLE{Kostogryz_2023,
       author = {{Kostogryz}, N. and {Shapiro}, A.~I. and {Witzke}, V. and {Grant}, D. and {Wakeford}, H.~R. and {Stevenson}, K.~B. and {Solanki}, S.~K. and {Gizon}, L.},
        title = "{MPS-ATLAS Library of Stellar Model Atmospheres and Spectra}",
      journal = {Research Notes of the American Astronomical Society},
         year = 2023,
        month = mar,
       volume = {7},
       number = {3},
          eid = {39},
        pages = {39},
          doi = {10.3847/2515-5172/acc180},
archivePrefix = {arXiv},
       eprint = {2303.02685},
 primaryClass = {astro-ph.SR},
       adsurl = {https://ui.adsabs.harvard.edu/abs/2023RNAAS...7...39K}
}

@INPROCEEDINGS{Castelli_2003,
       author = {{Castelli}, F. and {Kurucz}, R.~L.},
        title = "{New Grids of ATLAS9 Model Atmospheres}",
    booktitle = {Modelling of Stellar Atmospheres},
         year = 2003,
       editor = {{Piskunov}, N. and {Weiss}, W.~W. and {Gray}, D.~F.},
       series = {IAU Symposium},
       volume = {210},
        month = jan,
        pages = {A20},
          doi = {10.48550/arXiv.astro-ph/0405087},
archivePrefix = {arXiv},
       eprint = {astro-ph/0405087},
 primaryClass = {astro-ph},
       adsurl = {https://ui.adsabs.harvard.edu/abs/2003IAUS..210P.A20C}
}

@ARTICLE{Ekberg_1986,
       author = {{Ekberg}, U. and {Eriksson}, K. and {Gustafsson}, B.},
        title = "{Line blanketing in model atmospheres of carbon stars.}",
      journal = {\aap},
         year = 1986,
        month = oct,
       volume = {167},
        pages = {304-310},
       adsurl = {https://ui.adsabs.harvard.edu/abs/1986A&A...167..304E}
}

@ARTICLE{Anusha2021,
       author = {{Anusha}, L.~S. and {Shapiro}, A.~I. and {Witzke}, V. and {Cernetic}, M. and {Solanki}, S.~K. and {Gizon}, L.},
        title = "{Radiative Transfer with Opacity Distribution Functions: Application to Narrowband Filters}",
      journal = {\apjs},
         year = 2021,
        month = jul,
       volume = {255},
       number = {1},
          eid = {3},
        pages = {3},
          doi = {10.3847/1538-4365/abfb72},
archivePrefix = {arXiv},
       eprint = {2104.13661},
 primaryClass = {astro-ph.SR},
       adsurl = {https://ui.adsabs.harvard.edu/abs/2021ApJS..255....3A}
}

@ARTICLE{Cerentic2019,
       author = {{Cernetic}, M. and {Shapiro}, A.~I. and {Witzke}, V. and {Krivova}, N.~A. and {Solanki}, S.~K. and {Tagirov}, R.~V.},
        title = "{Opacity distribution functions for stellar spectra synthesis}",
      journal = {\aap},
         year = 2019,
        month = jul,
       volume = {627},
          eid = {A157},
        pages = {A157},
          doi = {10.1051/0004-6361/201935723},
archivePrefix = {arXiv},
       eprint = {1906.03112},
 primaryClass = {astro-ph.SR},
       adsurl = {https://ui.adsabs.harvard.edu/abs/2019A&A...627A.157C}
}

@ARTICLE{Goody1952,
       author = {{Goody}, R.~M.},
        title = "{A statistical model for water-vapour absorption}",
      journal = {Quarterly Journal of the Royal Meteorological Society},
         year = 1952,
        month = apr,
       volume = {78},
       number = {336},
        pages = {165-169},
          doi = {10.1002/qj.49707833604},
       adsurl = {https://ui.adsabs.harvard.edu/abs/1952QJRMS..78..165G}
}

@ARTICLE{Jorgensen2024,
       author = {{J{\o}rgensen}, Uffe G. and {Amadio}, Flavia and {Campos Estrada}, Beatriz and {M{\o}ller}, Kristian Holten and {Schneider}, Aaron D. and {Balduin}, Thorsten and {D'Alessandro}, Azzurra and {Symeonidou}, Eftychia and {Helling}, Christiane and {Nordlund}, {\r{A}}ke and {Woitke}, Peter},
        title = "{A grid of self-consistent MSG (MARCS-StaticWeather-GGchem) cool stellar, sub-stellar, and exoplanetary model atmospheres}",
      journal = {\aap},
         year = 2024,
        month = oct,
       volume = {690},
          eid = {A127},
        pages = {A127},
          doi = {10.1051/0004-6361/202450108},
archivePrefix = {arXiv},
       eprint = {2407.09397},
 primaryClass = {astro-ph.EP},
       adsurl = {https://ui.adsabs.harvard.edu/abs/2024A&A...690A.127J}
}

@ARTICLE{Gustafsson2008,
       author = {{Gustafsson}, B. and {Edvardsson}, B. and {Eriksson}, K. and {J{\o}rgensen}, U.~G. and {Nordlund}, {\r{A}}. and {Plez}, B.},
        title = "{A grid of MARCS model atmospheres for late-type stars. I. Methods and general properties}",
      journal = {\aap},
         year = 2008,
        month = aug,
       volume = {486},
       number = {3},
        pages = {951-970},
          doi = {10.1051/0004-6361:200809724},
archivePrefix = {arXiv},
       eprint = {0805.0554},
 primaryClass = {astro-ph},
       adsurl = {https://ui.adsabs.harvard.edu/abs/2008A&A...486..951G}
}

@ARTICLE{Malkmus1967,
       author = {{Malkmus}, W.},
        title = "{Random Lorentz Band Model with Exponential-Tailed S\^{\ensuremath{-}}1 Line-Intensity Distribution Function*}",
      journal = {Journal of the Optical Society of America (1917-1983)},
         year = 1967,
        month = mar,
       volume = {57},
       number = {3},
        pages = {323},
          doi = {10.1364/JOSA.57.000323},
       adsurl = {https://ui.adsabs.harvard.edu/abs/1967JOSA...57..323M}
}

@ARTICLE{Saxner1984,
       author = {{Saxner}, M. and {Gustafsson}, B.},
        title = "{A method for adding opacity distribution functions}",
      journal = {\aap},
         year = 1984,
        month = nov,
       volume = {140},
       number = {2},
        pages = {334-340},
       adsurl = {https://ui.adsabs.harvard.edu/abs/1984A&A...140..334S}
}

@ARTICLE{Kurucz1979,
       author = {{Kurucz}, R.~L.},
        title = "{Model atmospheres for G, F, A, B, and O stars.}",
      journal = {\apjs},
         year = 1979,
        month = may,
       volume = {40},
        pages = {1-340},
          doi = {10.1086/190589},
       adsurl = {https://ui.adsabs.harvard.edu/abs/1979ApJS...40....1K}
}

@ARTICLE{Labs1951,
       author = {{Labs}, D.},
        title = "{Die Temperaturschichtung der Sonnenatmosph{\"a}re im Strahlungsgleichgewicht unter Ber{\"u}cksichtigung der Fraunhoferlinien. Mit 6 Textabbildungen}",
      journal = {\zap},
         year = 1951,
        month = jan,
       volume = {29},
        pages = {199},
       adsurl = {https://ui.adsabs.harvard.edu/abs/1951ZA.....29..199L}
}

@ARTICLE{MacDonald2023,
       author = {{MacDonald}, Ryan J. and {Batalha}, Natasha E.},
        title = "{A Catalog of Exoplanet Atmospheric Retrieval Codes}",
      journal = {Research Notes of the American Astronomical Society},
         year = 2023,
        month = mar,
       volume = {7},
       number = {3},
          eid = {54},
        pages = {54},
          doi = {10.3847/2515-5172/acc46a},
archivePrefix = {arXiv},
       eprint = {2303.12925},
 primaryClass = {astro-ph.EP},
       adsurl = {https://ui.adsabs.harvard.edu/abs/2023RNAAS...7...54M}
}

@ARTICLE{Barstow2022,
       author = {{Barstow}, Joanna K. and {Changeat}, Quentin and {Chubb}, Katy L. and {Cubillos}, Patricio E. and {Edwards}, Billy and {MacDonald}, Ryan J. and {Min}, Michiel and {Waldmann}, Ingo P.},
        title = "{A retrieval challenge exercise for the Ariel mission}",
      journal = {Experimental Astronomy},
         year = 2022,
        month = apr,
       volume = {53},
       number = {2},
        pages = {447-471},
          doi = {10.1007/s10686-021-09821-w},
archivePrefix = {arXiv},
       eprint = {2203.00482},
 primaryClass = {astro-ph.EP},
       adsurl = {https://ui.adsabs.harvard.edu/abs/2022ExA....53..447B}
}

@ARTICLE{Malik2017,
       author = {{Malik}, Matej and {Grosheintz}, Luc and {Mendon{\c{c}}a}, Jo{\~a}o M. and {Grimm}, Simon L. and {Lavie}, Baptiste and {Kitzmann}, Daniel and {Tsai}, Shang-Min and {Burrows}, Adam and {Kreidberg}, Laura and {Bedell}, Megan and {Bean}, Jacob L. and {Stevenson}, Kevin B. and {Heng}, Kevin},
        title = "{HELIOS: An Open-source, GPU-accelerated Radiative Transfer Code for Self-consistent Exoplanetary Atmospheres}",
      journal = {\aj},
         year = 2017,
        month = feb,
       volume = {153},
       number = {2},
          eid = {56},
        pages = {56},
          doi = {10.3847/1538-3881/153/2/56},
archivePrefix = {arXiv},
       eprint = {1606.05474},
 primaryClass = {astro-ph.EP},
       adsurl = {https://ui.adsabs.harvard.edu/abs/2017AJ....153...56M}
}

@ARTICLE{Baudino2015,
       author = {{Baudino}, J. -L. and {B{\'e}zard}, B. and {Boccaletti}, A. and {Bonnefoy}, M. and {Lagrange}, A. -M. and {Galicher}, R.},
        title = "{Interpreting the photometry and spectroscopy of directly imaged planets: a new atmospheric model applied to {\ensuremath{\beta}} Pictoris b and SPHERE observations}",
      journal = {\aap},
         year = 2015,
        month = oct,
       volume = {582},
          eid = {A83},
        pages = {A83},
          doi = {10.1051/0004-6361/201526332},
archivePrefix = {arXiv},
       eprint = {1504.04876},
 primaryClass = {astro-ph.EP},
       adsurl = {https://ui.adsabs.harvard.edu/abs/2015A&A...582A..83B}
}

@ARTICLE{Phillips2020,
       author = {{Phillips}, M.~W. and {Tremblin}, P. and {Baraffe}, I. and {Chabrier}, G. and {Allard}, N.~F. and {Spiegelman}, F. and {Goyal}, J.~M. and {Drummond}, B. and {H{\'e}brard}, E.},
        title = "{A new set of atmosphere and evolution models for cool T-Y brown dwarfs and giant exoplanets}",
      journal = {\aap},
         year = 2020,
        month = may,
       volume = {637},
          eid = {A38},
        pages = {A38},
          doi = {10.1051/0004-6361/201937381},
archivePrefix = {arXiv},
       eprint = {2003.13717},
 primaryClass = {astro-ph.SR},
       adsurl = {https://ui.adsabs.harvard.edu/abs/2020A&A...637A..38P}
}

@ARTICLE{Mukherjee2024,
       author = {{Mukherjee}, Sagnick and {Fortney}, Jonathan J. and {Morley}, Caroline V. and {Batalha}, Natasha E. and {Marley}, Mark S. and {Karalidi}, Theodora and {Visscher}, Channon and {Lupu}, Roxana and {Freedman}, Richard and {Gharib-Nezhad}, Ehsan},
        title = "{The Sonora Substellar Atmosphere Models. IV. Elf Owl: Atmospheric Mixing and Chemical Disequilibrium with Varying Metallicity and C/O Ratios}",
      journal = {\apj},
         year = 2024,
        month = mar,
       volume = {963},
       number = {1},
          eid = {73},
        pages = {73},
          doi = {10.3847/1538-4357/ad18c2},
archivePrefix = {arXiv},
       eprint = {2402.00756},
 primaryClass = {astro-ph.EP},
       adsurl = {https://ui.adsabs.harvard.edu/abs/2024ApJ...963...73M}
}

@article{McKay1979,
 ISSN = {00401706},
 URL = {http://www.jstor.org/stable/1268522},
 author = {M. D. McKay and R. J. Beckman and W. J. Conover},
 journal = {Technometrics},
 number = {2},
 pages = {239--245},
 publisher = {[Taylor & Francis, Ltd., American Statistical Association, American Society for Quality]},
 title = {A Comparison of Three Methods for Selecting Values of Input Variables in the Analysis of Output from a Computer Code},
 urldate = {2025-08-09},
 volume = {21},
 year = {1979}
}

@ARTICLE{Hogan2020,
       author = {{Hogan}, Robin J. and {Matricardi}, Marco},
        title = "{Evaluating and improving the treatment of gases in radiation schemes: the Correlated K-Distribution Model Intercomparison Project (CKDMIP)}",
      journal = {Geoscientific Model Development},
         year = 2020,
        month = dec,
       volume = {13},
       number = {12},
        pages = {6501-6521},
          doi = {10.5194/gmd-13-6501-2020},
       adsurl = {https://ui.adsabs.harvard.edu/abs/2020GMD....13.6501H}
}

@BOOK{Pierrehumbert2010,
       author = {{Pierrehumbert}, Raymond T.},
        title = "{Principles of Planetary Climate}",
         year = 2010,
publisher = "{Cambridge University Press}",
       adsurl = {https://ui.adsabs.harvard.edu/abs/2010ppc..book.....P}
}

@ARTICLE{Min2020,
       author = {{Min}, Michiel and {Ormel}, Chris W. and {Chubb}, Katy and {Helling}, Christiane and {Kawashima}, Yui},
        title = "{The ARCiS framework for exoplanet atmospheres. Modelling philosophy and retrieval}",
      journal = {\aap},
         year = 2020,
        month = oct,
       volume = {642},
          eid = {A28},
        pages = {A28},
          doi = {10.1051/0004-6361/201937377},
archivePrefix = {arXiv},
       eprint = {2006.12821},
 primaryClass = {astro-ph.EP},
       adsurl = {https://ui.adsabs.harvard.edu/abs/2020A&A...642A..28M}
}

@BOOK{Bracewell1965,
       author = {{Bracewell}, Ron},
        title = "{The Fourier Transform and its applications}",
         year = 1965,
        publisher = "{McGraw-Hill Electrical and Electronic Engineering Series, New York: McGraw-Hill}",
       adsurl = {https://ui.adsabs.harvard.edu/abs/1965ftia.book.....B}
}

@ARTICLE{Baudino2017,
       author = {{Baudino}, Jean-Loup and {Molli{\`e}re}, Paul and {Venot}, Olivia and {Tremblin}, Pascal and {B{\'e}zard}, Bruno and {Lagage}, Pierre-Olivier},
        title = "{Toward the Analysis of JWST  Exoplanet Spectra: Identifying Troublesome Model Parameters}",
      journal = {\apj},
         year = 2017,
        month = dec,
       volume = {850},
       number = {2},
          eid = {150},
        pages = {150},
          doi = {10.3847/1538-4357/aa95be},
archivePrefix = {arXiv},
       eprint = {1710.08235},
 primaryClass = {astro-ph.EP},
       adsurl = {https://ui.adsabs.harvard.edu/abs/2017ApJ...850..150B}
}

@ARTICLE{Nixon2024,
       author = {{Nixon}, Matthew C. and {Welbanks}, Luis and {McGill}, Peter and {Kempton}, Eliza M. -R.},
        title = "{Methods for Incorporating Model Uncertainty into Exoplanet Atmospheric Analysis}",
      journal = {\apj},
         year = 2024,
        month = may,
       volume = {966},
       number = {2},
          eid = {156},
        pages = {156},
          doi = {10.3847/1538-4357/ad354e},
archivePrefix = {arXiv},
       eprint = {2310.03713},
 primaryClass = {astro-ph.EP},
       adsurl = {https://ui.adsabs.harvard.edu/abs/2024ApJ...966..156N}
}

@ARTICLE{Matthews2025,
       author = {{Matthews}, Elisabeth C. and {Molli{\`e}re}, Paul and {K{\"u}hnle}, Helena and {Patapis}, Polychronis and {Whiteford}, Niall and {Samland}, Matthias and {Lagage}, Pierre-Olivier and {Waters}, Rens and {Tsai}, Shang-Min and {Zahnle}, Kevin and {Guedel}, Manuel and {Henning}, Thomas and {Vandenbussche}, Bart and {Absil}, Olivier and {Argyriou}, Ioannis and {Barrado}, David and {Coulais}, Alain and {Glauser}, Adrian M. and {Olofsson}, Goran and {Pye}, John P. and {Rouan}, Daniel and {Royer}, Pierre and {van Dishoeck}, Ewine F. and {Ray}, T.~P. and {{\"O}stlin}, G{\"o}ran},
        title = "{HCN and C$_{2}$H$_{2}$ in the Atmosphere of a T8.5+T9 Brown Dwarf Binary}",
      journal = {\apjl},
         year = 2025,
        month = mar,
       volume = {981},
       number = {2},
          eid = {L31},
        pages = {L31},
          doi = {10.3847/2041-8213/adb4ec},
archivePrefix = {arXiv},
       eprint = {2502.13610},
 primaryClass = {astro-ph.EP},
       adsurl = {https://ui.adsabs.harvard.edu/abs/2025ApJ...981L..31M}
}

@ARTICLE{McCarthy2025,
       author = {{McCarthy}, Allison M. and {Vos}, Johanna M. and {Muirhead}, Philip S. and {Biller}, Beth A. and {Morley}, Caroline V. and {Faherty}, Jacqueline and {Burningham}, Ben and {Calamari}, Emily and {Cowan}, Nicolas B. and {Cruz}, Kelle L. and {Gonzales}, Eileen and {Limbach}, Mary Anne and {Liu}, Pengyu and {Nasedkin}, Evert and {Su{\'a}rez}, Genaro and {Tan}, Xianyu and {O'Toole}, Cian and {Visscher}, Channon and {Whiteford}, Niall and {Zhou}, Yifan},
        title = "{The JWST Weather Report from the Isolated Exoplanet Analog SIMP 0136+0933: Pressure-dependent Variability Driven by Multiple Mechanisms}",
      journal = {\apjl},
         year = 2025,
        month = mar,
       volume = {981},
       number = {2},
          eid = {L22},
        pages = {L22},
          doi = {10.3847/2041-8213/ad9eaf},
archivePrefix = {arXiv},
       eprint = {2411.16577},
 primaryClass = {astro-ph.EP},
       adsurl = {https://ui.adsabs.harvard.edu/abs/2025ApJ...981L..22M}
}

@ARTICLE{Biller2024,
       author = {{Biller}, Beth A. and {Vos}, Johanna M. and {Zhou}, Yifan and {McCarthy}, Allison M. and {Tan}, Xianyu and {Crossfield}, Ian J.~M. and {Whiteford}, Niall and {Suarez}, Genaro and {Faherty}, Jacqueline and {Manjavacas}, Elena and {Chen}, Xueqing and {Liu}, Pengyu and {Sutlieff}, Ben J. and {Limbach}, Mary Anne and {Molliere}, Paul and {Dupuy}, Trent J. and {Oliveros-Gomez}, Natalia and {Muirhead}, Philip S. and {Henning}, Thomas and {Mace}, Gregory and {Crouzet}, Nicolas and {Karalidi}, Theodora and {Morley}, Caroline V. and {Tremblin}, Pascal and {Kataria}, Tiffany},
        title = "{The JWST weather report from the nearest brown dwarfs I: multiperiod JWST NIRSpec + MIRI monitoring of the benchmark binary brown dwarf WISE 1049AB}",
      journal = {\mnras},
         year = 2024,
        month = aug,
       volume = {532},
       number = {2},
        pages = {2207-2233},
          doi = {10.1093/mnras/stae1602},
archivePrefix = {arXiv},
       eprint = {2407.09194},
 primaryClass = {astro-ph.SR},
       adsurl = {https://ui.adsabs.harvard.edu/abs/2024MNRAS.532.2207B}
}

@ARTICLE{Miles2023,
       author = {{Miles}, Brittany E. and {Biller}, Beth A. and {Patapis}, Polychronis and {Worthen}, Kadin and {Rickman}, Emily and {Hoch}, Kielan K.~W. and {Skemer}, Andrew and {Perrin}, Marshall D. and {Whiteford}, Niall and {Chen}, Christine H. and {Sargent}, B. and {Mukherjee}, Sagnick and {Morley}, Caroline V. and {Moran}, Sarah E. and {Bonnefoy}, Mickael and {Petrus}, Simon and {Carter}, Aarynn L. and {Choquet}, Elodie and {Hinkley}, Sasha and {Ward-Duong}, Kimberly and {Leisenring}, Jarron M. and {Millar-Blanchaer}, Maxwell A. and {Pueyo}, Laurent and {Ray}, Shrishmoy and {Sallum}, Steph and {Stapelfeldt}, Karl R. and {Stone}, Jordan M. and {Wang}, Jason J. and {Absil}, Olivier and {Balmer}, William O. and {Boccaletti}, Anthony and {Bonavita}, Mariangela and {Booth}, Mark and {Bowler}, Brendan P. and {Chauvin}, Gael and {Christiaens}, Valentin and {Currie}, Thayne and {Danielski}, Camilla and {Fortney}, Jonathan J. and {Girard}, Julien H. and {Grady}, Carol A. and {Greenbaum}, Alexandra Z. and {Henning}, Thomas and {Hines}, Dean C. and {Janson}, Markus and {Kalas}, Paul and {Kammerer}, Jens and {Kennedy}, Grant M. and {Kenworthy}, Matthew A. and {Kervella}, Pierre and {Lagage}, Pierre-Olivier and {Lew}, Ben W.~P. and {Liu}, Michael C. and {Macintosh}, Bruce and {Marino}, Sebastian and {Marley}, Mark S. and {Marois}, Christian and {Matthews}, Elisabeth C. and {Matthews}, Brenda C. and {Mawet}, Dimitri and {McElwain}, Michael W. and {Metchev}, Stanimir and {Meyer}, Michael R. and {Molliere}, Paul and {Pantin}, Eric and {Quirrenbach}, Andreas and {Rebollido}, Isabel and {Ren}, Bin B. and {Schneider}, Glenn and {Vasist}, Malavika and {Wyatt}, Mark C. and {Zhou}, Yifan and {Briesemeister}, Zackery W. and {Bryan}, Marta L. and {Calissendorff}, Per and {Cantalloube}, Faustine and {Cugno}, Gabriele and {De Furio}, Matthew and {Dupuy}, Trent J. and {Factor}, Samuel M. and {Faherty}, Jacqueline K. and {Fitzgerald}, Michael P. and {Franson}, Kyle and {Gonzales}, Eileen C. and {Hood}, Callie E. and {Howe}, Alex R. and {Kraus}, Adam L. and {Kuzuhara}, Masayuki and {Lagrange}, Anne-Marie and {Lawson}, Kellen and {Lazzoni}, Cecilia and {Liu}, Pengyu and {Llop-Sayson}, Jorge and {Lloyd}, James P. and {Martinez}, Raquel A. and {Mazoyer}, Johan and {Quanz}, Sascha P. and {Redai}, Jea Adams and {Samland}, Matthias and {Schlieder}, Joshua E. and {Tamura}, Motohide and {Tan}, Xianyu and {Uyama}, Taichi and {Vigan}, Arthur and {Vos}, Johanna M. and {Wagner}, Kevin and {Wolff}, Schuyler G. and {Ygouf}, Marie and {Zhang}, Xi and {Zhang}, Keming and {Zhang}, Zhoujian},
        title = "{The JWST Early-release Science Program for Direct Observations of Exoplanetary Systems II: A 1 to 20 {\ensuremath{\mu}}m Spectrum of the Planetary-mass Companion VHS 1256-1257 b}",
      journal = {\apjl},
         year = 2023,
        month = mar,
       volume = {946},
       number = {1},
          eid = {L6},
        pages = {L6},
          doi = {10.3847/2041-8213/acb04a},
archivePrefix = {arXiv},
       eprint = {2209.00620},
 primaryClass = {astro-ph.EP},
       adsurl = {https://ui.adsabs.harvard.edu/abs/2023ApJ...946L...6M}
}

@ARTICLE{Malik2019,
       author = {{Malik}, Matej and {Kitzmann}, Daniel and {Mendon{\c{c}}a}, Jo{\~a}o M. and {Grimm}, Simon L. and {Marleau}, Gabriel-Dominique and {Linder}, Esther F. and {Tsai}, Shang-Min and {Heng}, Kevin},
        title = "{Self-luminous and Irradiated Exoplanetary Atmospheres Explored with HELIOS}",
      journal = {\aj},
         year = 2019,
        month = may,
       volume = {157},
       number = {5},
          eid = {170},
        pages = {170},
          doi = {10.3847/1538-3881/ab1084},
archivePrefix = {arXiv},
       eprint = {1903.06794},
 primaryClass = {astro-ph.EP},
       adsurl = {https://ui.adsabs.harvard.edu/abs/2019AJ....157..170M}
}

@ARTICLE{Mukherjee2023,
       author = {{Mukherjee}, Sagnick and {Batalha}, Natasha E. and {Fortney}, Jonathan J. and {Marley}, Mark S.},
        title = "{PICASO 3.0: A One-dimensional Climate Model for Giant Planets and Brown Dwarfs}",
      journal = {\apj},
         year = 2023,
        month = jan,
       volume = {942},
       number = {2},
          eid = {71},
        pages = {71},
          doi = {10.3847/1538-4357/ac9f48},
archivePrefix = {arXiv},
       eprint = {2208.07836},
 primaryClass = {astro-ph.EP},
       adsurl = {https://ui.adsabs.harvard.edu/abs/2023ApJ...942...71M}
}

@ARTICLE{Agrawal2024,
       author = {{Agrawal}, Arnav and {MacDonald}, Ryan},
        title = "{Cthulhu: An Open Source Molecular and Atomic Cross Section Computation Code for Substellar Atmospheres}",
      journal = {The Journal of Open Source Software},
         year = 2024,
        month = oct,
       volume = {9},
       number = {102},
          eid = {6894},
        pages = {6894},
          doi = {10.21105/joss.06894},
archivePrefix = {arXiv},
       eprint = {2410.14751},
 primaryClass = {astro-ph.IM},
       adsurl = {https://ui.adsabs.harvard.edu/abs/2024JOSS....9.6894A}
}

@ARTICLE{Yurchenko2018,
       author = {{Yurchenko}, Sergei N. and {Al-Refaie}, Ahmed F. and {Tennyson}, Jonathan},
        title = "{EXOCROSS: a general program for generating spectra from molecular line lists}",
      journal = {\aap},
         year = 2018,
        month = jun,
       volume = {614},
          eid = {A131},
        pages = {A131},
          doi = {10.1051/0004-6361/201732531},
archivePrefix = {arXiv},
       eprint = {1801.09803},
 primaryClass = {astro-ph.EP},
       adsurl = {https://ui.adsabs.harvard.edu/abs/2018A&A...614A.131Y}
}

@ARTICLE{Marley2021,
       author = {{Marley}, Mark S. and {Saumon}, Didier and {Visscher}, Channon and {Lupu}, Roxana and {Freedman}, Richard and {Morley}, Caroline and {Fortney}, Jonathan J. and {Seay}, Christopher and {Smith}, Adam J.~R.~W. and {Teal}, D.~J. and {Wang}, Ruoyan},
        title = "{The Sonora Brown Dwarf Atmosphere and Evolution Models. I. Model Description and Application to Cloudless Atmospheres in Rainout Chemical Equilibrium}",
      journal = {\apj},
         year = 2021,
        month = oct,
       volume = {920},
       number = {2},
          eid = {85},
        pages = {85},
          doi = {10.3847/1538-4357/ac141d},
archivePrefix = {arXiv},
       eprint = {2107.07434},
 primaryClass = {astro-ph.SR},
       adsurl = {https://ui.adsabs.harvard.edu/abs/2021ApJ...920...85M}
}

@ARTICLE{Edwards1996,
       author = {{Edwards}, J.~M.},
        title = "{Efficient Calculation of Infrared Fluxes and Cooling Rates Using the Two-Stream Equations.}",
      journal = {Journal of the Atmospheric Sciences},
         year = 1996,
        month = jul,
       volume = {53},
       number = {13},
        pages = {1921-1932},
          doi = {10.1175/1520-0469(1996)053<1921:ECOIFA>2.0.CO;2},
       adsurl = {https://ui.adsabs.harvard.edu/abs/1996JAtS...53.1921E}
}

@ARTICLE{Lee2024,
       author = {{Lee}, Elspeth K.~H. and {Tan}, Xianyu and {Tsai}, Shang-Min},
        title = "{Dynamically coupled kinetic chemistry in brown dwarf atmospheres - II. Cloud and chemistry connections in directly imaged sub-Jupiter exoplanets}",
      journal = {\mnras},
         year = 2024,
        month = apr,
       volume = {529},
       number = {3},
        pages = {2686-2701},
          doi = {10.1093/mnras/stae537},
archivePrefix = {arXiv},
       eprint = {2311.16722},
 primaryClass = {astro-ph.EP},
       adsurl = {https://ui.adsabs.harvard.edu/abs/2024MNRAS.529.2686L}
}

@ARTICLE{Zamyatina2024,
       author = {{Zamyatina}, Maria and {Christie}, Duncan A. and {H{\'e}brard}, Eric and {Mayne}, Nathan J. and {Radica}, Michael and {Taylor}, Jake and {Baskett}, Harry and {Moore}, Ben and {Lils}, Craig and {Sergeev}, Denis E. and {Ahrer}, Eva-Maria and {Manners}, James and {Kohary}, Krisztian and {Feinstein}, Adina D.},
        title = "{Quenching-driven equatorial depletion and limb asymmetries in hot Jupiter atmospheres: WASP-96b example}",
      journal = {\mnras},
         year = 2024,
        month = apr,
       volume = {529},
       number = {2},
        pages = {1776-1801},
          doi = {10.1093/mnras/stae600},
archivePrefix = {arXiv},
       eprint = {2402.14535},
 primaryClass = {astro-ph.EP},
       adsurl = {https://ui.adsabs.harvard.edu/abs/2024MNRAS.529.1776Z}
}

@ARTICLE{Drummond2020,
       author = {{Drummond}, Benjamin and {H{\'e}brard}, Eric and {Mayne}, Nathan J. and {Venot}, Olivia and {Ridgway}, Robert J. and {Changeat}, Quentin and {Tsai}, Shang-Min and {Manners}, James and {Tremblin}, Pascal and {Abraham}, Nathan Luke and {Sing}, David and {Kohary}, Krisztian},
        title = "{Implications of three-dimensional chemical transport in hot Jupiter atmospheres: Results from a consistently coupled chemistry-radiation-hydrodynamics model}",
      journal = {\aap},
         year = 2020,
        month = apr,
       volume = {636},
          eid = {A68},
        pages = {A68},
          doi = {10.1051/0004-6361/201937153},
archivePrefix = {arXiv},
       eprint = {2001.11444},
 primaryClass = {astro-ph.EP},
       adsurl = {https://ui.adsabs.harvard.edu/abs/2020A&A...636A..68D}
}

@ARTICLE{Barstow2020,
       author = {{Barstow}, Joanna K. and {Changeat}, Quentin and {Garland}, Ryan and {Line}, Michael R. and {Rocchetto}, Marco and {Waldmann}, Ingo P.},
        title = "{A comparison of exoplanet spectroscopic retrieval tools}",
      journal = {\mnras},
         year = 2020,
        month = apr,
       volume = {493},
       number = {4},
        pages = {4884-4909},
          doi = {10.1093/mnras/staa548},
archivePrefix = {arXiv},
       eprint = {2002.01063},
 primaryClass = {astro-ph.EP},
       adsurl = {https://ui.adsabs.harvard.edu/abs/2020MNRAS.493.4884B}
}

@software{Line2025,
       author = {{Line}, M.~R. and {Lustig-Yaeger}, J. and {Batalha}, N. and {Marley}, M. and {Zhang}, X. and {Wolf}, A.},
        title = "{CHIMERA: CaltecH Inverse ModEling and Retrieval Algorithms}",
 howpublished = {Astrophysics Source Code Library, record ascl:2504.014},
         year = 2025,
        month = apr,
          eid = {ascl:2504.014},
       adsurl = {https://ui.adsabs.harvard.edu/abs/2025ascl.soft04014L}
}

@ARTICLE{Batalha2019,
       author = {{Batalha}, Natasha E. and {Marley}, Mark S. and {Lewis}, Nikole K. and {Fortney}, Jonathan J.},
        title = "{Exoplanet Reflected-light Spectroscopy with PICASO}",
      journal = {\apj},
         year = 2019,
        month = jun,
       volume = {878},
       number = {1},
          eid = {70},
        pages = {70},
          doi = {10.3847/1538-4357/ab1b51},
archivePrefix = {arXiv},
       eprint = {1904.09355},
 primaryClass = {astro-ph.EP},
       adsurl = {https://ui.adsabs.harvard.edu/abs/2019ApJ...878...70B}
}

@ARTICLE{Leconte2021,
       author = {{Leconte}, J{\'e}r{\'e}my},
        title = "{Spectral binning of precomputed correlated-k coefficients}",
      journal = {\aap},
         year = 2021,
        month = jan,
       volume = {645},
          eid = {A20},
        pages = {A20},
          doi = {10.1051/0004-6361/202039040},
archivePrefix = {arXiv},
       eprint = {2012.01428},
 primaryClass = {astro-ph.EP},
       adsurl = {https://ui.adsabs.harvard.edu/abs/2021A&A...645A..20L}
}

@ARTICLE{Garland2019,
       author = {{Garland}, R. and {Irwin}, P.~G.~J.},
        title = "{Effectively Calculating Gaseous Absorption in Radiative Transfer Models of Exoplanetary and Brown Dwarf Atmospheres}",
      journal = {arXiv e-prints},
         year = 2019,
        month = mar,
          eid = {arXiv:1903.03997},
        pages = {arXiv:1903.03997},
          doi = {10.48550/arXiv.1903.03997},
archivePrefix = {arXiv},
       eprint = {1903.03997},
 primaryClass = {astro-ph.EP},
       adsurl = {https://ui.adsabs.harvard.edu/abs/2019arXiv190303997G}
}

@ARTICLE{Woitke2018,
       author = {{Woitke}, P. and {Helling}, Ch. and {Hunter}, G.~H. and {Millard}, J.~D. and {Turner}, G.~E. and {Worters}, M. and {Blecic}, J. and {Stock}, J.~W.},
        title = "{Equilibrium chemistry down to 100 K. Impact of silicates and phyllosilicates on the carbon to oxygen ratio}",
      journal = {\aap},
         year = 2018,
        month = jun,
       volume = {614},
          eid = {A1},
        pages = {A1},
          doi = {10.1051/0004-6361/201732193},
archivePrefix = {arXiv},
       eprint = {1712.01010},
 primaryClass = {astro-ph.EP},
       adsurl = {https://ui.adsabs.harvard.edu/abs/2018A&A...614A...1W}
}

@ARTICLE{Amundsen2014,
       author = {{Amundsen}, David S. and {Baraffe}, Isabelle and {Tremblin}, Pascal and {Manners}, James and {Hayek}, Wolfgang and {Mayne}, Nathan J. and {Acreman}, David M.},
        title = "{Accuracy tests of radiation schemes used in hot Jupiter global circulation models}",
      journal = {\aap},
         year = 2014,
        month = apr,
       volume = {564},
          eid = {A59},
        pages = {A59},
          doi = {10.1051/0004-6361/201323169},
archivePrefix = {arXiv},
       eprint = {1402.0814},
 primaryClass = {astro-ph.EP},
       adsurl = {https://ui.adsabs.harvard.edu/abs/2014A&A...564A..59A}
}

@ARTICLE{Deitrick2022,
       author = {{Deitrick}, Russell and {Heng}, Kevin and {Schroffenegger}, Urs and {Kitzmann}, Daniel and {Grimm}, Simon L. and {Malik}, Matej and {Mendon{\c{c}}a}, Jo{\~a}o M. and {Morris}, Brett M.},
        title = "{The THOR + HELIOS general circulation model: multiwavelength radiative transfer with accurate scattering by clouds/hazes}",
      journal = {\mnras},
         year = 2022,
        month = may,
       volume = {512},
       number = {3},
        pages = {3759-3787},
          doi = {10.1093/mnras/stac680},
archivePrefix = {arXiv},
       eprint = {2203.02293},
 primaryClass = {astro-ph.EP},
       adsurl = {https://ui.adsabs.harvard.edu/abs/2022MNRAS.512.3759D}
}

@ARTICLE{Lee2021,
       author = {{Lee}, Elspeth K.~H. and {Parmentier}, Vivien and {Hammond}, Mark and {Grimm}, Simon L. and {Kitzmann}, Daniel and {Tan}, Xianyu and {Tsai}, Shang-Min and {Pierrehumbert}, Raymond T.},
        title = "{Simulating gas giant exoplanet atmospheres with EXO-FMS: comparing semigrey, picket fence, and correlated-k radiative-transfer schemes}",
      journal = {\mnras},
         year = 2021,
        month = sep,
       volume = {506},
       number = {2},
        pages = {2695-2711},
          doi = {10.1093/mnras/stab1851},
archivePrefix = {arXiv},
       eprint = {2106.11664},
 primaryClass = {astro-ph.EP},
       adsurl = {https://ui.adsabs.harvard.edu/abs/2021MNRAS.506.2695L}
}

@ARTICLE{Grimm2015,
       author = {{Grimm}, Simon L. and {Heng}, Kevin},
        title = "{HELIOS-K: An Ultrafast, Open-source Opacity Calculator for Radiative Transfer}",
      journal = {\apj},
         year = 2015,
        month = aug,
       volume = {808},
       number = {2},
          eid = {182},
        pages = {182},
          doi = {10.1088/0004-637X/808/2/182},
archivePrefix = {arXiv},
       eprint = {1503.03806},
 primaryClass = {astro-ph.EP},
       adsurl = {https://ui.adsabs.harvard.edu/abs/2015ApJ...808..182G}
}

@ARTICLE{Lacis1991,
       author = {{Lacis}, A.~A. and {Oinas}, V.},
        title = "{A description of the correlated-k distribution method for modelling nongray gaseous absorption, thermal emission, and multiple scattering in vertically inhomogeneous atmospheres}",
      journal = {\jgr},
         year = 1991,
        month = may,
       volume = {96},
        pages = {9027-9064},
          doi = {10.1029/90JD01945},
       adsurl = {https://ui.adsabs.harvard.edu/abs/1991JGR....96.9027L}
}

@ARTICLE{Fu1992,
       author = {{Fu}, Qiang and {Liou}, K.~N.},
        title = "{On the correlated k-distribution method for radiative transfer in nonhomogeneous atmospheres}",
      journal = {Journal of the Atmospheric Sciences},
         year = 1992,
        month = nov,
       volume = {49},
       number = {22},
        pages = {2139-2156},
          doi = {10.1175/1520-0469(1992)049<2139:OTCDMF>2.0.CO;2},
       adsurl = {https://ui.adsabs.harvard.edu/abs/1992JAtS...49.2139F}
}

@ARTICLE{Goody1989,
       author = {{Goody}, Richard and {West}, Robert and {Chen}, Luke and {Crisp}, David},
        title = "{The correlated-k method for radiation calculations in nonhomogeneous atmospheres.}",
      journal = {\jqsrt},
         year = 1989,
        month = dec,
       volume = {42},
        pages = {539-550},
          doi = {10.1016/0022-4073(89)90044-7},
       adsurl = {https://ui.adsabs.harvard.edu/abs/1989JQSRT..42..539G}
}

@ARTICLE{Tsai2021,
       author = {{Tsai}, Shang-Min and {Malik}, Matej and {Kitzmann}, Daniel and {Lyons}, James R. and {Fateev}, Alexander and {Lee}, Elspeth and {Heng}, Kevin},
        title = "{A Comparative Study of Atmospheric Chemistry with VULCAN}",
      journal = {\apj},
         year = 2021,
        month = dec,
       volume = {923},
       number = {2},
          eid = {264},
        pages = {264},
          doi = {10.3847/1538-4357/ac29bc},
archivePrefix = {arXiv},
       eprint = {2108.01790},
 primaryClass = {astro-ph.EP},
       adsurl = {https://ui.adsabs.harvard.edu/abs/2021ApJ...923..264T}
}

@article{Hogan2024,
author = {Hogan, Robin J.},
title = {What are the optimum discrete angles to use in thermal-infrared radiative transfer calculations?},
journal = {Quarterly Journal of the Royal Meteorological Society},
volume = {150},
number = {758},
pages = {318-333},
doi = {https://doi.org/10.1002/qj.4598},
url = {https://rmets.onlinelibrary.wiley.com/doi/abs/10.1002/qj.4598},
eprint = {https://rmets.onlinelibrary.wiley.com/doi/pdf/10.1002/qj.4598},
year = {2024}
}

@ARTICLE{Olson1987,
       author = {{Olson}, G.~L. and {Kunasz}, P.~B.},
        title = "{Short characteristic solution of the non-LTE transfer problem by operator perturbation. I. The one-dimensional planar slab.}",
      journal = {\jqsrt},
         year = 1987,
        month = jan,
       volume = {38},
       number = {5},
        pages = {325-336},
          doi = {10.1016/0022-4073(87)90027-6},
       adsurl = {https://ui.adsabs.harvard.edu/abs/1987JQSRT..38..325O}
}

@ARTICLE{Addison2019,
       author = {{Addison}, Brett and {Wright}, Duncan J. and {Wittenmyer}, Robert A. and {Horner}, Jonathan and {Mengel}, Matthew W. and {Johns}, Daniel and {Marti}, Connor and {Nicholson}, Belinda and {Soutter}, Jack and {Bowler}, Brendan and {Crossfield}, Ian and {Kane}, Stephen R. and {Kielkopf}, John and {Plavchan}, Peter and {Tinney}, C.~G. and {Zhang}, Hui and {Clark}, Jake T. and {Clerte}, Mathieu and {Eastman}, Jason D. and {Swift}, Jon and {Bottom}, Michael and {Muirhead}, Philip and {McCrady}, Nate and {Herzig}, Erich and {Hogstrom}, Kristina and {Wilson}, Maurice and {Sliski}, David and {Johnson}, Samson A. and {Wright}, Jason T. and {Johnson}, John Asher and {Blake}, Cullen and {Riddle}, Reed and {Lin}, Brian and {Cornachione}, Matthew and {Bedding}, Timothy R. and {Stello}, Dennis and {Huber}, Daniel and {Marsden}, Stephen and {Carter}, Bradley D.},
        title = "{Minerva-Australis. I. Design, Commissioning, and First Photometric Results}",
      journal = {\pasp},
         year = 2019,
        month = nov,
       volume = {131},
       number = {1005},
        pages = {115003},
          doi = {10.1088/1538-3873/ab03aa},
archivePrefix = {arXiv},
       eprint = {1901.11231},
 primaryClass = {astro-ph.IM},
       adsurl = {https://ui.adsabs.harvard.edu/abs/2019PASP..131k5003A}
}

@ARTICLE{Parmentier2015,
       author = {{Parmentier}, Vivien and {Guillot}, Tristan and {Fortney}, Jonathan J. and {Marley}, Mark S.},
        title = "{A non-grey analytical model for irradiated atmospheres. II. Analytical vs. numerical solutions}",
      journal = {\aap},
         year = 2015,
        month = feb,
       volume = {574},
          eid = {A35},
        pages = {A35},
          doi = {10.1051/0004-6361/201323127},
archivePrefix = {arXiv},
       eprint = {1311.6322},
 primaryClass = {astro-ph.EP},
       adsurl = {https://ui.adsabs.harvard.edu/abs/2015A&A...574A..35P}
}

@ARTICLE{Parmentier2014,
       author = {{Parmentier}, Vivien and {Guillot}, Tristan},
        title = "{A non-grey analytical model for irradiated atmospheres. I. Derivation}",
      journal = {\aap},
         year = 2014,
        month = feb,
       volume = {562},
          eid = {A133},
        pages = {A133},
          doi = {10.1051/0004-6361/201322342},
archivePrefix = {arXiv},
       eprint = {1311.6597},
 primaryClass = {astro-ph.EP},
       adsurl = {https://ui.adsabs.harvard.edu/abs/2014A&A...562A.133P}
}

@ARTICLE{Stock2018,
       author = {{Stock}, Joachim W. and {Kitzmann}, Daniel and {Patzer}, A. Beate C. and {Sedlmayr}, Erwin},
        title = "{FastChem: A computer program for efficient complex chemical equilibrium calculations in the neutral/ionized gas phase with applications to stellar and planetary atmospheres}",
      journal = {\mnras},
         year = 2018,
        month = sep,
       volume = {479},
       number = {1},
        pages = {865-874},
          doi = {10.1093/mnras/sty1531},
archivePrefix = {arXiv},
       eprint = {1804.05010},
 primaryClass = {astro-ph.EP},
       adsurl = {https://ui.adsabs.harvard.edu/abs/2018MNRAS.479..865S}
}

@ARTICLE{Showman2009,
       author = {{Showman}, Adam P. and {Fortney}, Jonathan J. and {Lian}, Yuan and {Marley}, Mark S. and {Freedman}, Richard S. and {Knutson}, Heather A. and {Charbonneau}, David},
        title = "{Atmospheric Circulation of Hot Jupiters: Coupled Radiative-Dynamical General Circulation Model Simulations of HD 189733b and HD 209458b}",
      journal = {\apj},
         year = 2009,
        month = jul,
       volume = {699},
       number = {1},
        pages = {564-584},
          doi = {10.1088/0004-637X/699/1/564},
archivePrefix = {arXiv},
       eprint = {0809.2089},
 primaryClass = {astro-ph},
       adsurl = {https://ui.adsabs.harvard.edu/abs/2009ApJ...699..564S}
}

@ARTICLE{Irwin2008,
       author = {{Irwin}, P.~G.~J. and {Teanby}, N.~A. and {de Kok}, R. and {Fletcher}, L.~N. and {Howett}, C.~J.~A. and {Tsang}, C.~C.~C. and {Wilson}, C.~F. and {Calcutt}, S.~B. and {Nixon}, C.~A. and {Parrish}, P.~D.},
        title = "{The NEMESIS planetary atmosphere radiative transfer and retrieval tool}",
      journal = {\jqsrt},
         year = 2008,
        month = apr,
       volume = {109},
        pages = {1136-1150},
          doi = {10.1016/j.jqsrt.2007.11.006},
       adsurl = {https://ui.adsabs.harvard.edu/abs/2008JQSRT.109.1136I}
}

@ARTICLE{Molliere2019,
       author = {{Molli{\`e}re}, P. and {Wardenier}, J.~P. and {van Boekel}, R. and {Henning}, Th. and {Molaverdikhani}, K. and {Snellen}, I.~A.~G.},
        title = "{petitRADTRANS. A Python radiative transfer package for exoplanet characterization and retrieval}",
      journal = {\aap},
         year = 2019,
        month = jul,
       volume = {627},
          eid = {A67},
        pages = {A67},
          doi = {10.1051/0004-6361/201935470},
archivePrefix = {arXiv},
       eprint = {1904.11504},
 primaryClass = {astro-ph.EP},
       adsurl = {https://ui.adsabs.harvard.edu/abs/2019A&A...627A..67M}
}

@ARTICLE{Schneider2024,
       author = {{Schneider}, Aaron David and {Molli{\`e}re}, Paul and {Louppe}, Gilles and {Carone}, Ludmila and {J{\o}rgensen}, Uffe Gr{\r{a}}e and {Decin}, Leen and {Helling}, Christiane},
        title = "{Harnessing machine learning for accurate treatment of overlapping opacity species in general circulation models}",
      journal = {\aap},
         year = 2024,
        month = feb,
       volume = {682},
          eid = {A79},
        pages = {A79},
          doi = {10.1051/0004-6361/202348221},
archivePrefix = {arXiv},
       eprint = {2311.00775},
 primaryClass = {astro-ph.EP},
       adsurl = {https://ui.adsabs.harvard.edu/abs/2024A&A...682A..79S}
}

@ARTICLE{Schneider2022,
       author = {{Schneider}, Aaron David and {Carone}, Ludmila and {Decin}, Leen and {J{\o}rgensen}, Uffe Gr{\r{a}}e and {Molli{\`e}re}, Paul and {Baeyens}, Robin and {Kiefer}, Sven and {Helling}, Christiane},
        title = "{Exploring the deep atmospheres of HD 209458b and WASP-43b using a non-gray general circulation model}",
      journal = {\aap},
         year = 2022,
        month = aug,
       volume = {664},
          eid = {A56},
        pages = {A56},
          doi = {10.1051/0004-6361/202142728},
archivePrefix = {arXiv},
       eprint = {2202.09183},
 primaryClass = {astro-ph.EP},
       adsurl = {https://ui.adsabs.harvard.edu/abs/2022A&A...664A..56S}
}

@ARTICLE{Tsai2022,
       author = {{Tsai}, Shang-Min and {Lee}, Elspeth K.~H. and {Pierrehumbert}, Raymond},
        title = "{A mini-chemical scheme with net reactions for 3D general circulation models. I. Thermochemical kinetics}",
      journal = {\aap},
         year = 2022,
        month = aug,
       volume = {664},
          eid = {A82},
        pages = {A82},
          doi = {10.1051/0004-6361/202142816},
archivePrefix = {arXiv},
       eprint = {2204.04201},
 primaryClass = {astro-ph.EP},
       adsurl = {https://ui.adsabs.harvard.edu/abs/2022A&A...664A..82T}
}

@ARTICLE{Kataria2013,
       author = {{Kataria}, T. and {Showman}, A.~P. and {Lewis}, N.~K. and {Fortney}, J.~J. and {Marley}, M.~S. and {Freedman}, R.~S.},
        title = "{Three-dimensional Atmospheric Circulation of Hot Jupiters on Highly Eccentric Orbits}",
      journal = {\apj},
         year = 2013,
        month = apr,
       volume = {767},
       number = {1},
          eid = {76},
        pages = {76},
          doi = {10.1088/0004-637X/767/1/76},
archivePrefix = {arXiv},
       eprint = {1208.3795},
 primaryClass = {astro-ph.EP},
       adsurl = {https://ui.adsabs.harvard.edu/abs/2013ApJ...767...76K}
}

@ARTICLE{Amundsen2017,
       author = {{Amundsen}, David S. and {Tremblin}, Pascal and {Manners}, James and {Baraffe}, Isabelle and {Mayne}, Nathan J.},
        title = "{Treatment of overlapping gaseous absorption with the correlated-k method in hot Jupiter and brown dwarf atmosphere models}",
      journal = {\aap},
         year = 2017,
        month = feb,
       volume = {598},
          eid = {A97},
        pages = {A97},
          doi = {10.1051/0004-6361/201629322},
archivePrefix = {arXiv},
       eprint = {1610.01389},
 primaryClass = {astro-ph.EP},
       adsurl = {https://ui.adsabs.harvard.edu/abs/2017A&A...598A..97A}
}

\begin{appendix}

\section{PRAS PDF convolution limiting case}
\label{app:PRAS_lim}

In this Appendix, I show that the Monte Carlo sampling method proposed in the main text, polynomial reconstruction and sampling (PRAS), is mathematically equivalent to convolution of the probability distribution function (PDF) of all species that are mixed, preserving the required properties for random overlap.
\citet{Saxner1984, Goody1989} and \citet{Amundsen2017} show that the random overlap of opacities is equivalent to convolution of individual species PDFs.

The PDF of two functions, representing the gas phase cross sections of gas $a$ and $b$ are given by $f_{a}$(k) and $f_{b}$(k). 
The convolution of both functions is given by the convolution integral \citep[e.g.][]{Bracewell1965}
\begin{equation}
\label{eq:A1}
  f_{\rm mix}(k) \approx (f_{a} * f_{b})(k) = \int_{0}^{k} f_{a}(k') f_{b}(k - k') \, dk.
\end{equation}
Given the inverse, normalised CDF of each PDF function in $g$ $\in$ [0,1] space, $k_{a}(g)$ = $F_{a}^{-1}$($g$) and $k_{b}(g)$ = $F_{b}^{-1}$($g$), sampling the CDF randomly and independently for each gas in $g$ for an individual sample $j$ and adding gives
\begin{equation}
    k_{\rm ab}^{(j)} = k_{a}(g_{j}) + k_{b}(g'_{j}).
\end{equation}
In the case of PRAS, $k_{a}(g)$ and $k_{b}(g)$ are represented by polynomial or spline functions to the cross-section CDF.
From the Monte Carlo principle of CDF sampling, sampling of $k_{a}(g_{j})$ and $k_{b}(g'_{j})$ results in a stochastic reconstruction of the PDFs of each species.
Through addition of independent samples, the convolved PDF of both species is produced, corresponding to the opacity distribution under the assumption of random overlap (i.e., the distribution of the sum of independent opacity samples from each species).
Following the principles of Monte Carlo integration, an integral can be estimated from sampling a number of independent random draws of the function to be integrated
\begin{equation}
    I = \int_{b}^{a} f(x) dx = \frac{b-a}{n_{\rm s}}\sum^{n_{\rm s}}_{j=1}f(\chi_{j}),
\end{equation}
where $f(\chi_{i})$ is a randomly drawn value of the underlying function $f(x)$.
The Monte Carlo estimator for the convolved PDF (Eq. \ref{eq:A1}) is then given as
\begin{equation}
\label{eq:A3}
      (f_{a} * f_{b})(k) \approx \frac{1}{n_{s}}\sum_{j=1}^{n_{s}}\delta\left[k - k_{a}(g_{j}) + k_{b}(g'_{j})\right] = \frac{1}{n_{s}}\sum_{j=1}^{n_{s}}\delta\left(k - k_{\rm ab}^{(j)}\right),
\end{equation}
where $n_{s}$ is the number of individual, independent samples, $j$, of $g$ and $g'$.
Here, $\delta\left(k - k_{ab}^{(j)}\right)$ is the Dirac delta function, which is a generalised function (distribution) that is zero everywhere except at \( k = k_{ab}^{(j)} \), and satisfies the identity
\begin{equation}
    \int_{-\infty}^{\infty} \delta(k - k')\,dk = 1.
\end{equation}
In this context, the delta function represents a unit probability located at the sampled opacity value $k_{ab}^{(j)}$. 
The summation in Eq. \eqref{eq:A3} therefore corresponds to an approximation of the convolution of the two opacity PDFs through Monte Carlo sampling. 
Each delta function contributes a spike at a sampled value from the random overlap of the two species' cross sections. 
As the number of samples $n_{s}$ increases
\begin{equation}
f_{\text{mix}}(k) \approx \frac{1}{n_{s}} \sum_{j=1}^{n_{s}} \delta(k - k_{ab}^{(j)}),
\end{equation}
converges (in distribution) to the random-overlapped, true convolved PDF of the mixture, i.e., $f_{a} * f_{b}$.
This is a highly useful property, as the practitioner can control the accuracy of the convolution through changing the number of samples and choice of random number generator.

For multiple gases, the convolution equation is extended to multiple dimensions, resulting in an expression with multiple integrals.
For brevity, I write the total convolution over species $1, \dots, n_{\rm sp}$ as
\begin{equation}
    f_{\rm mix}(k) = (f_{a} * f_{b} * f_{c} \dots f_{n_{\rm sp}})(k),
\end{equation}
where $n_{\rm sp}$ is the number of species to mix.
However, the Monte Carlo estimator easily generalises to multiple species through
\begin{equation}
      f_{\rm mix}(k) \approx \frac{1}{n_{s}}\sum_{j=1}^{n_{s}}\delta\left(k - \sum_{i=1}^{n_{\rm sp}}k_{i}^{(j)}\right),
\end{equation}
where the $\sum_{n=i}^{n_{\rm sp}}k_{i}^{(j)}$ is the independently sampled CDF of each species.
In this way, the multidimensional integral is sampled and all species mixed together in one step.

\section{k-coefficient recovery at extreme T-p}
\label{app:mix_ex}

\begin{figure*}
    \centering
    \includegraphics[width=0.40\linewidth]{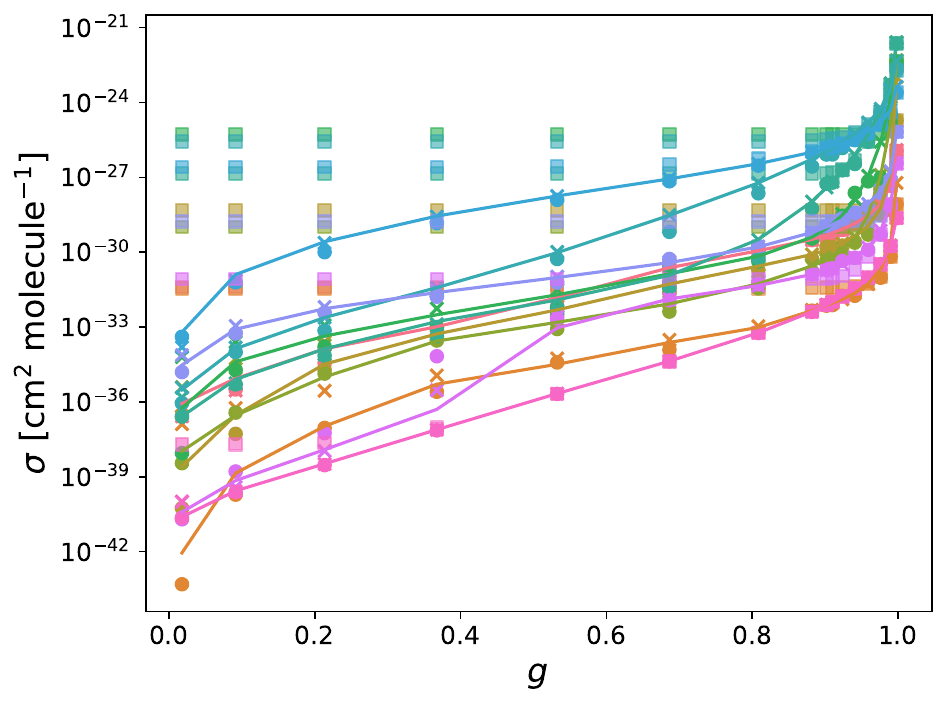}
    \includegraphics[width=0.40\linewidth]{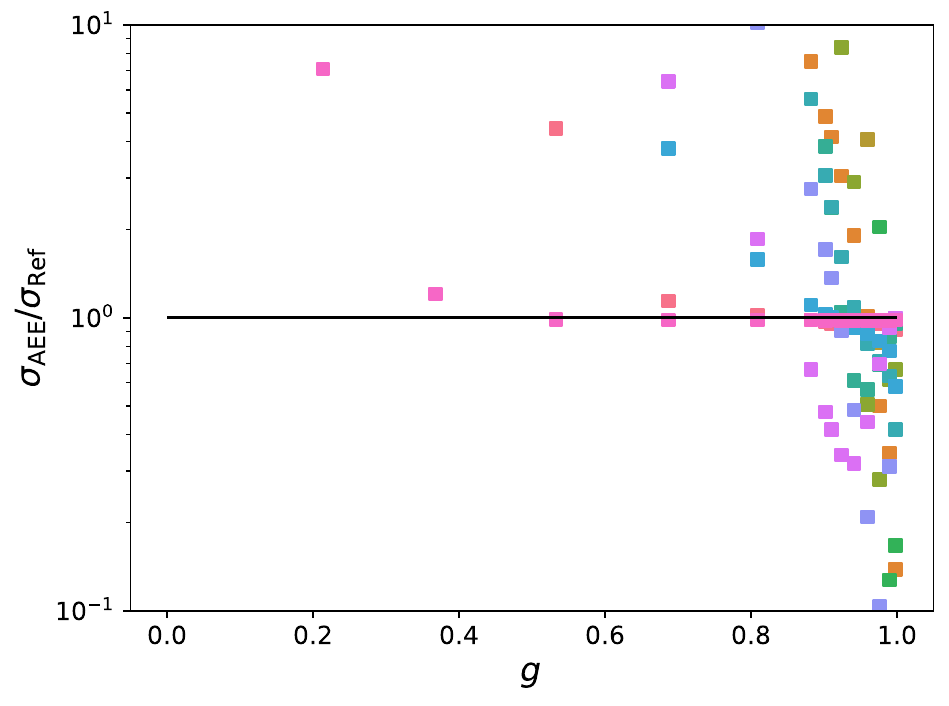}
    \includegraphics[width=0.40\linewidth]{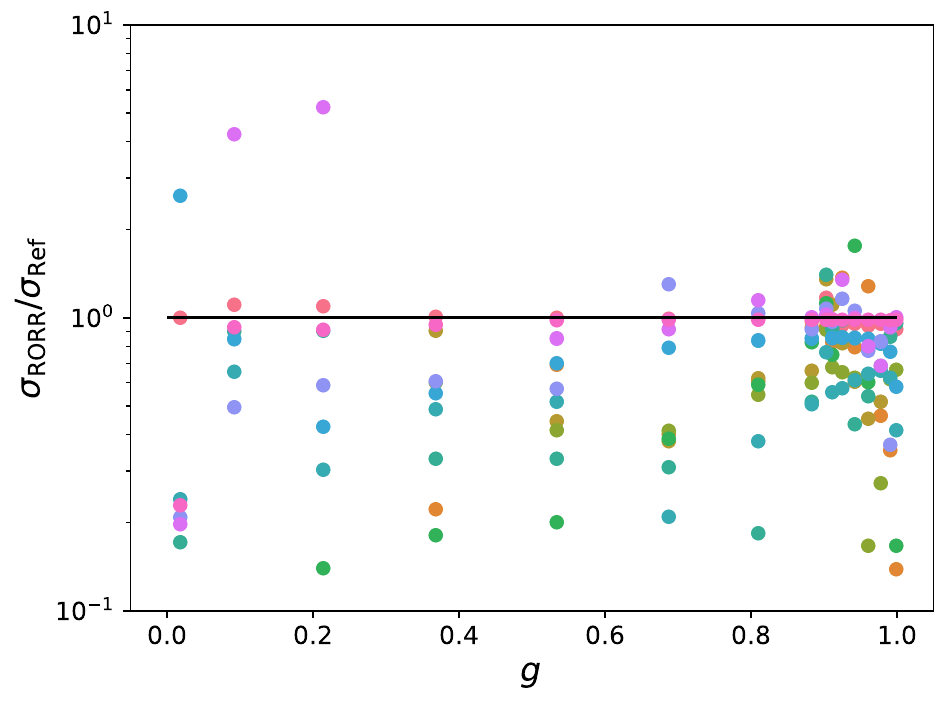}
    \includegraphics[width=0.40\linewidth]{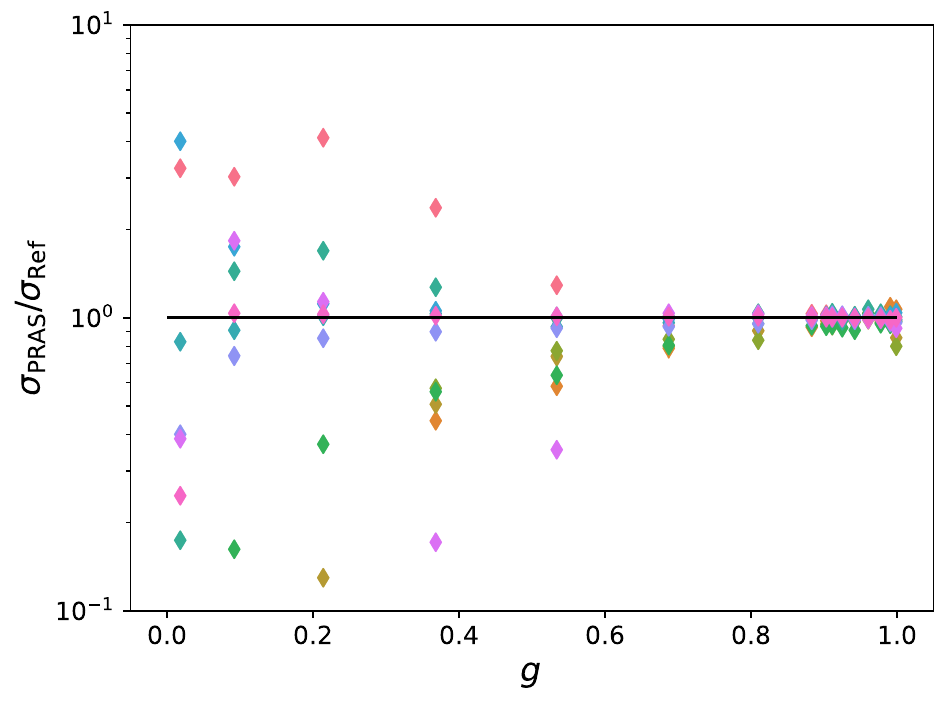}
    \caption{Comparison between methods at $T$ = 50 K, $p$ = 10$^{-8}$ bar.}
    \label{app:mix_ex_1}
\end{figure*}

\begin{figure*}
    \centering
    \includegraphics[width=0.40\linewidth]{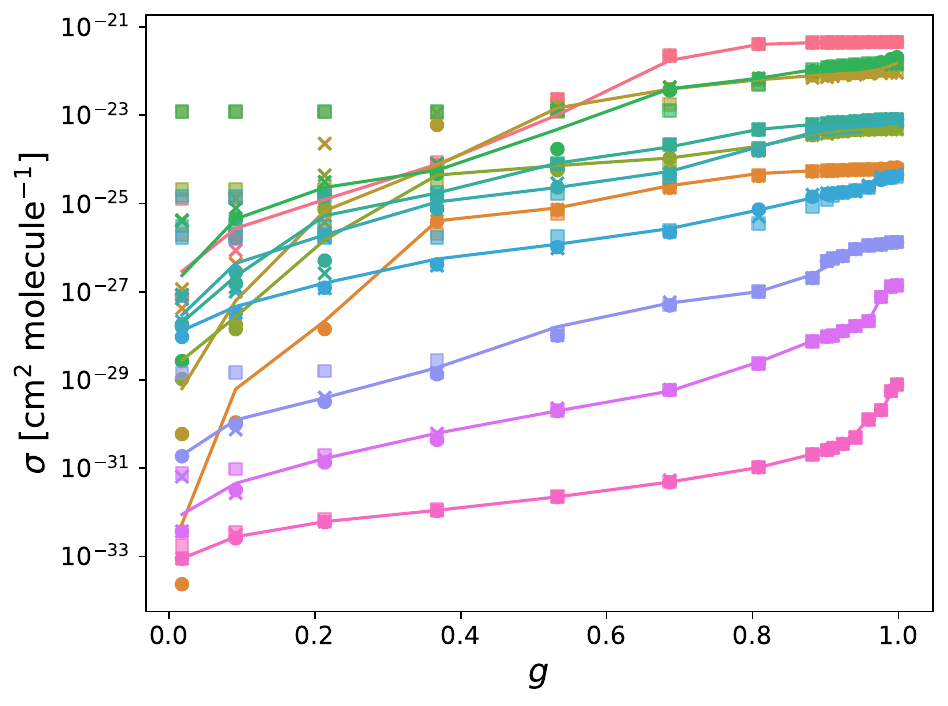}
    \includegraphics[width=0.40\linewidth]{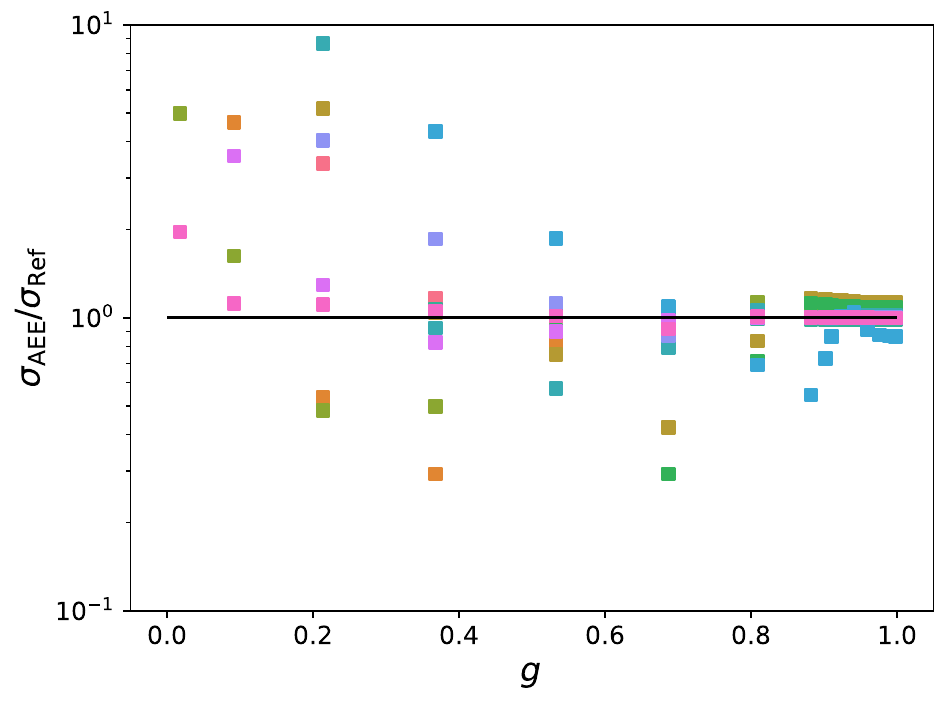}
    \includegraphics[width=0.40\linewidth]{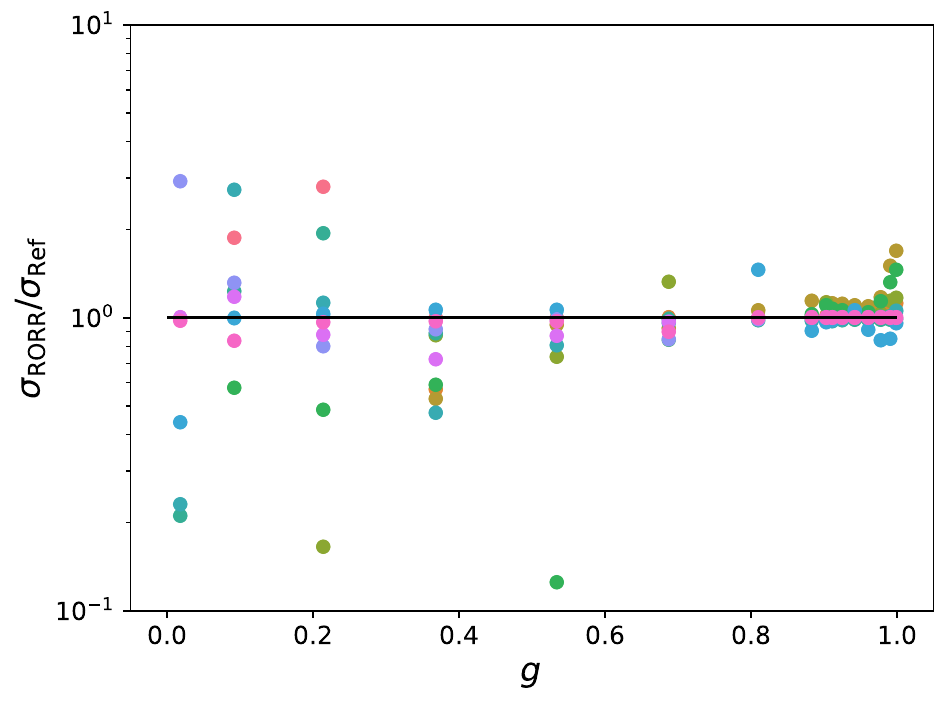}
    \includegraphics[width=0.40\linewidth]{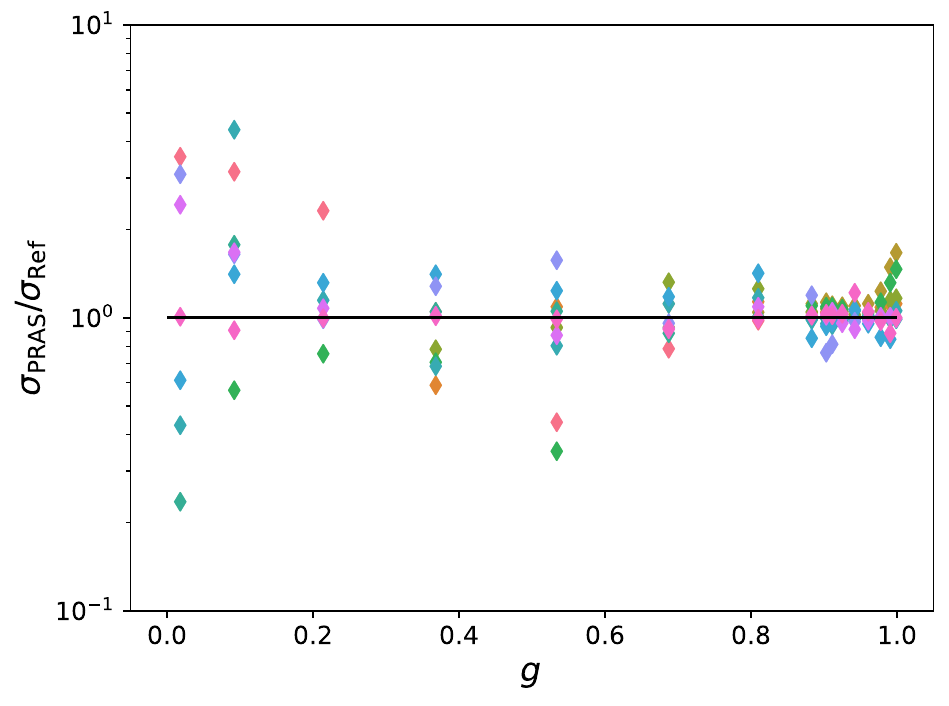}
    \caption{Comparison between methods at $T$ = 50 K, $p$ = 1000 bar.}
    \label{app:mix_ex_2}
\end{figure*}

\begin{figure*}
    \centering
    \includegraphics[width=0.40\linewidth]{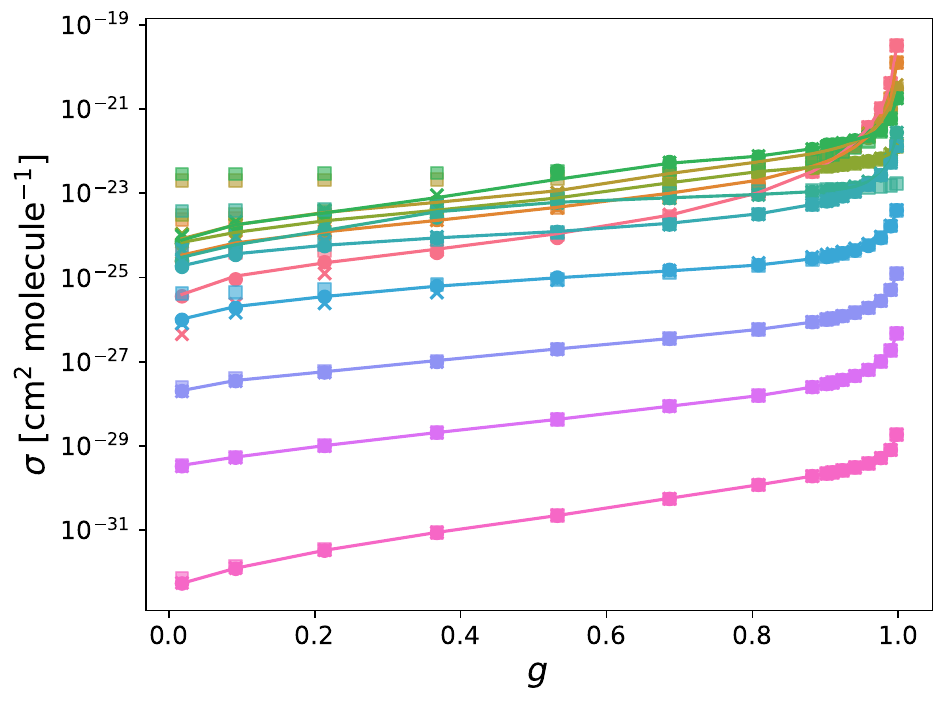}
    \includegraphics[width=0.40\linewidth]{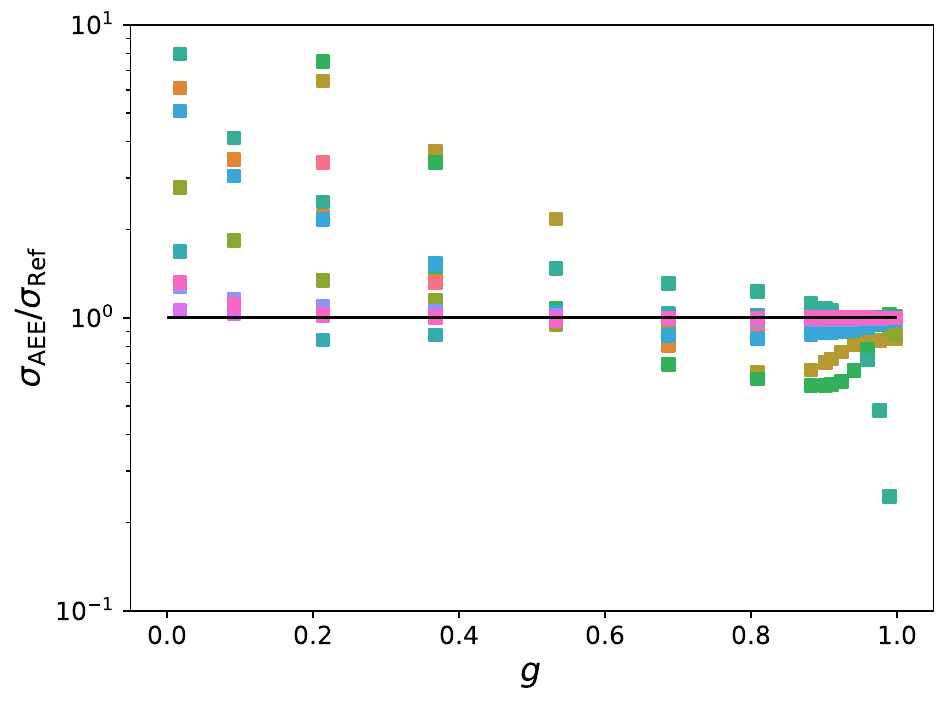}
    \includegraphics[width=0.40\linewidth]{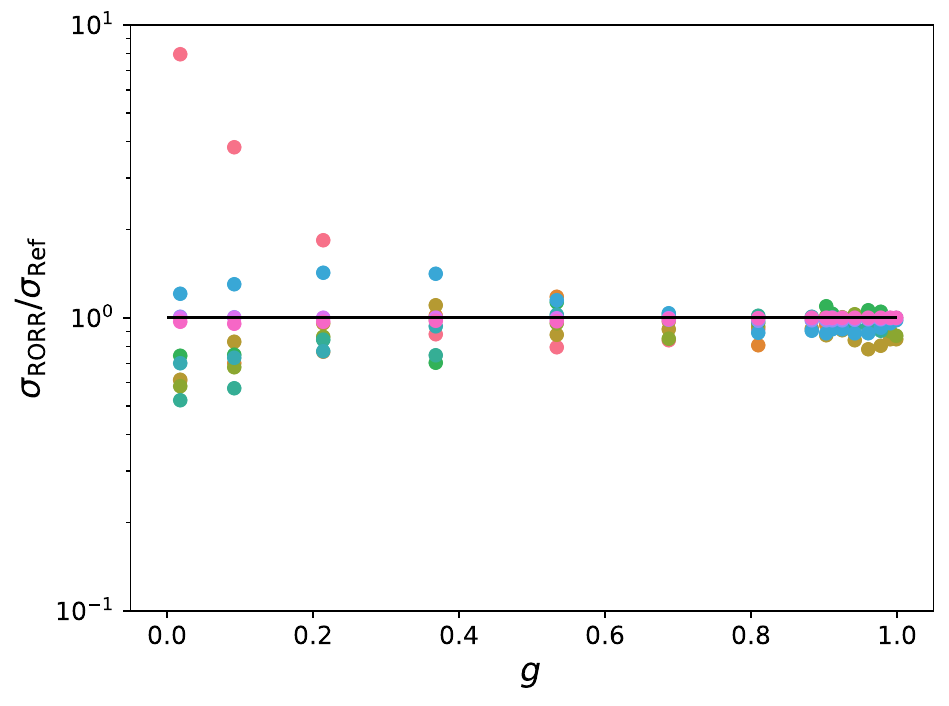}
    \includegraphics[width=0.40\linewidth]{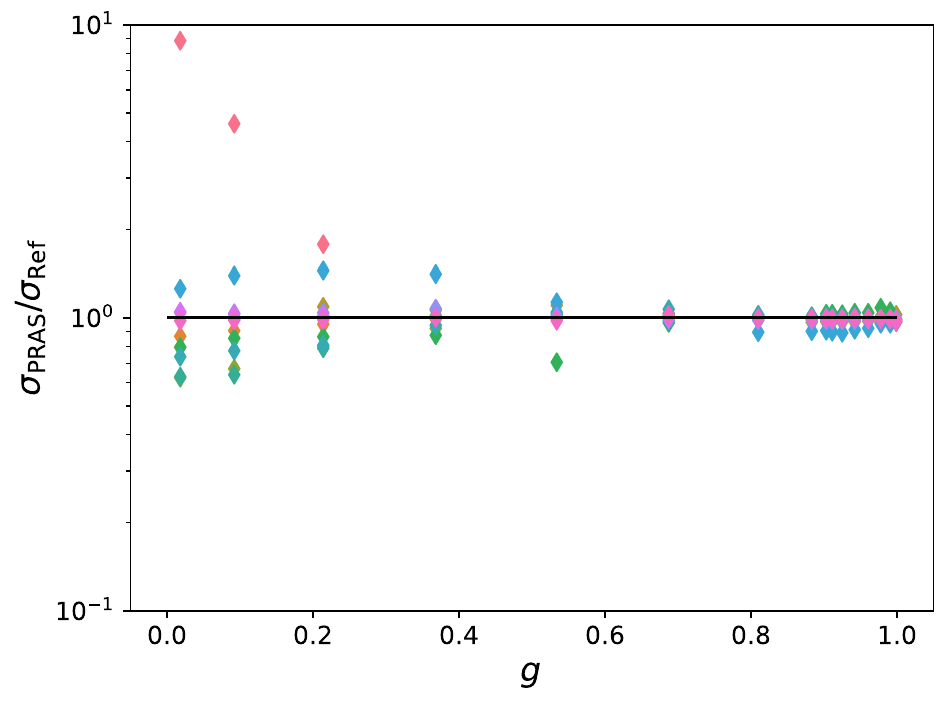}
    \caption{Comparison between methods at $T$ = 1900 K, $p$ = 10$^{-8}$ bar.}
    \label{app:mix_ex_3}
\end{figure*}

\begin{figure*}
    \centering
    \includegraphics[width=0.40\linewidth]{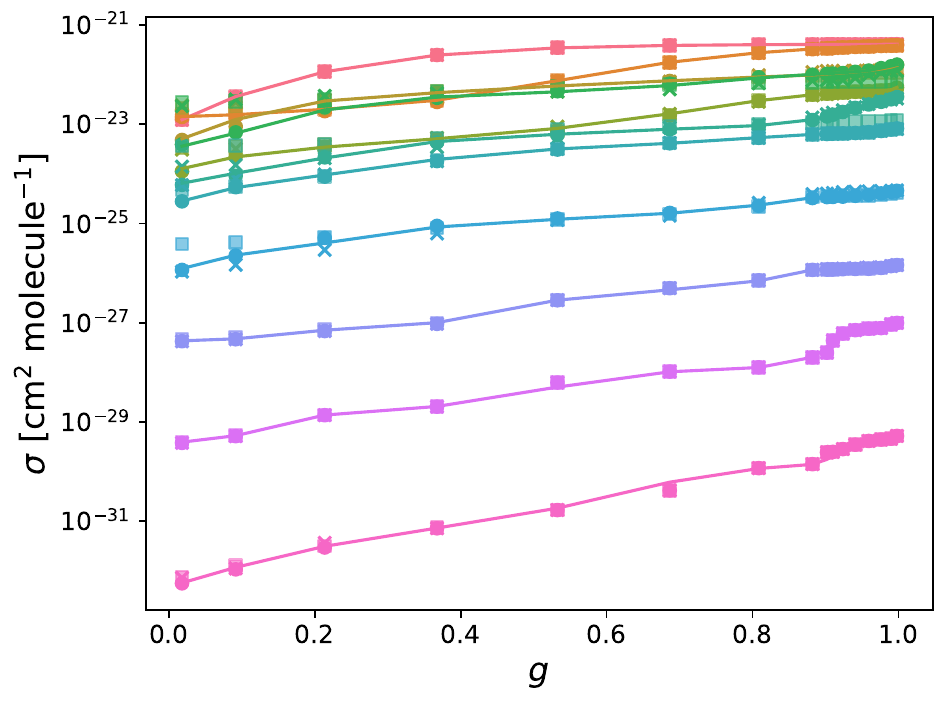}
    \includegraphics[width=0.40\linewidth]{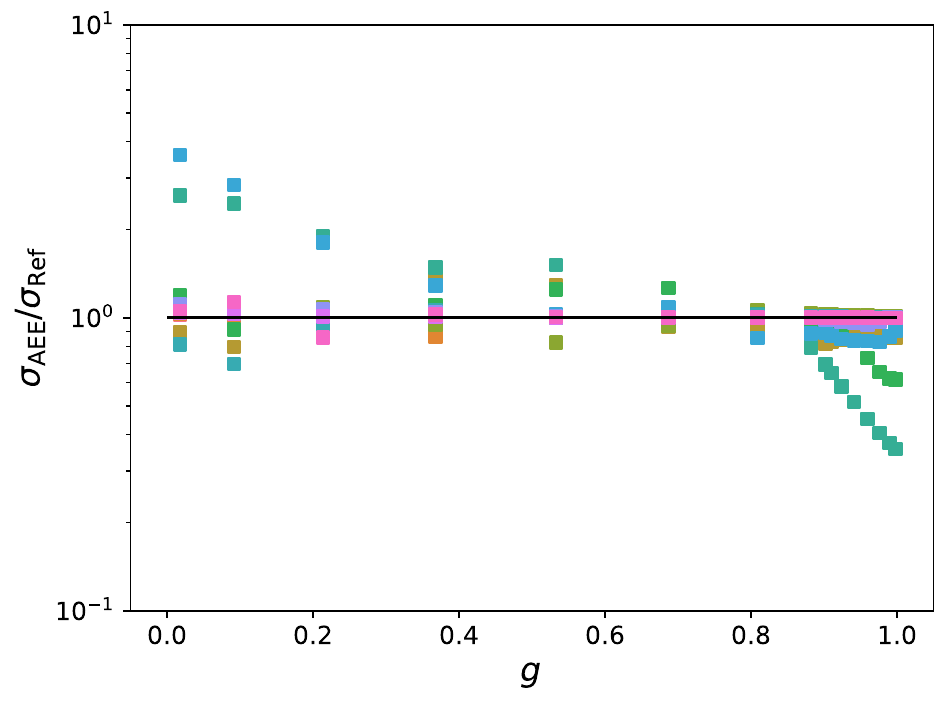}
    \includegraphics[width=0.40\linewidth]{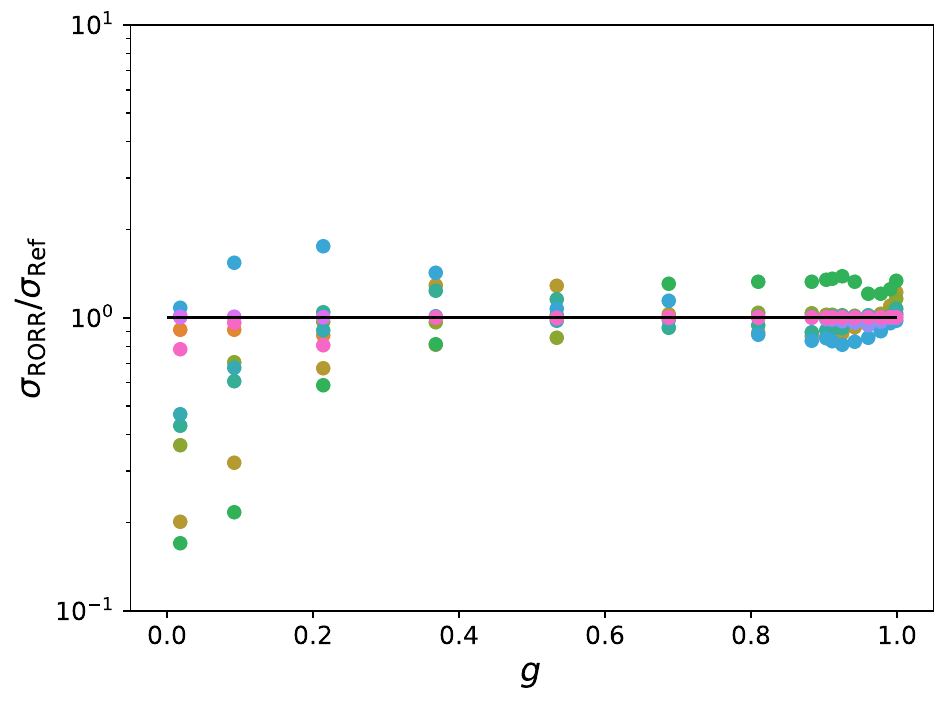}
    \includegraphics[width=0.40\linewidth]{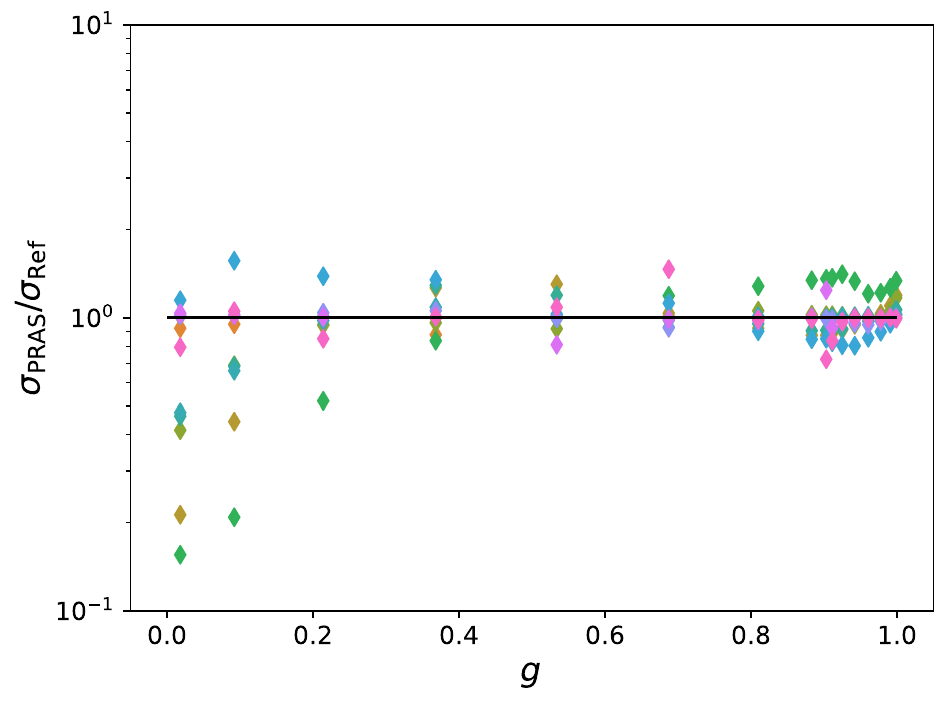}
    \caption{Comparison between methods at $T$ = 1900 K, $p$ = 1000 bar.}
    \label{app:mix_ex_4}
\end{figure*}

In this Appendix, I produce plots in the style of Figure \ref{fig:mix_ex} but for four T-p points that correspond to the extreme T-p edges of the available original cross-section data for each species: $T$ = 50 K, $p$ = 10$^{-8}$ bar (Figure \ref{app:mix_ex_1}), $T$ = 50 K, $p$ = 1000 bar (Figure \ref{app:mix_ex_2}), $T$ = 1900 K, $p$ = 10$^{-8}$ bar (Figure \ref{app:mix_ex_3}) and $T$ = 1900 K, $p$ = 1000 bar (Figure \ref{app:mix_ex_4}).

\end{appendix}

\end{document}